\documentclass[%
 reprint,
 amsmath,amssymb,
 aps,nofootinbib
]{revtex4-2}

\usepackage{graphicx}
\usepackage{dcolumn}
\usepackage{bm}
\usepackage{hyperref}

\usepackage{tabularx}

\graphicspath{
	{plots/},{figures/}, {TC/},{fits/}} 
\usepackage{subcaption}
\usepackage{comment}
\usepackage{float}
\usepackage{xcolor}
\usepackage[normalem]{ulem}
\usepackage{color}
\definecolor{darkred}{rgb}{0.4,0.0,0.0}
\definecolor{darkbred}{rgb}{0.7,0.0,0.0}
\definecolor{darkgreen}{rgb}{0.0,0.4,0.0}
\definecolor{darkblue}{rgb}{0.0,0.0,0.4}
\definecolor{darkmagenta}{rgb}{0.55, 0.0, 0.55}
\definecolor{bole}{rgb}{0.47, 0.27, 0.23}

\newcommand{\be}{\begin{equation}}
\newcommand{\ee}{\end{equation}}
\newcommand{\beq}{\begin{equation}}
\newcommand{\eeq}{\end{equation}}
\newcommand{\bea}{\begin{eqnarray}}
\newcommand{\eea}{\end{eqnarray}}

\newcommand{\bie}{\begin{itemize}}

\newcommand{\eie}{\end{itemize} }
\newcommand{\ben}{ \begin{enumerate}}
\newcommand{\een}{\end{enumerate} }
\newcommand{\bbl}{\begin{block}{} \begin{center}}
\newcommand{\ebl}{\end{center} \end{block}}
\newcommand{\bref}{\begin{flushright} \begin{tiny}}
\newcommand{\eref}{\end{tiny} \end{flushright}}

\newcommand{\Tr}{\text{Tr}}

\begin{document}

\preprint{APS/123-QED}

\title{The spectrum of open confining strings in the large-$N_c$ limit.}

\author{Alireza Sharifian}
\email{alireza.sharifian@tecnico.ulisboa.pt}
\affiliation{CeFEMA, Departamento de F\'isica, Instituto Superior T\'ecnico (Universidade  de Lisboa),
Avenida Rovisco Pais, 1049-001 Lisboa, Portugal}
\author{Andreas Athenodorou}
\email{a.athenodorou@cyi.ac.cy}
\affiliation{Computation-based Science and Technology Research Center, The Cyprus Institute, 20 Kavafi Str., Nicosia 2121, Cyprus}
\author{Pedro Bicudo}%
\email{bicudo@tecnico.ulisboa.pt}
\affiliation{CeFEMA, Departamento de F\'isica, Instituto Superior T\'ecnico (Universidade de Lisboa),
Avenida Rovisco Pais, ${\it 1049}$-001 Lisboa, Portugal}


\date{\today}

\begin{abstract}
{\color{black} In this study, we conduct a thorough examination of the spectrum of the open confining string in 3+1 dimensions, commonly referred to as the open flux-tube, across various gauge groups of $SU(N_c)$. Our primary objective is to explore its behaviour as we approach the large-$N_c$ limit and the identification of possible world-sheet axion states. Specifically, we undertake a detailed analysis of the associated spectrum for $N_c=3, 4, 5, 6$. This marks the first systematic investigation of the open flux-tube spectrum within the context of the large-$N_c$ limit. More specifically, we analyse the spectra of flux-tubes that form between a static quark-antiquark pair, considering a significant number of radial excitations and eight irreducible representations characterized by the quantum numbers of angular momentum $\Lambda$, charge conjugation and parity $\eta_{CP}$ and the reflection symmetry $\epsilon$ for $\Lambda=0$. To this purpose we employ a diverse set of suitable operators, an anisotropic action, smearing techniques, and solve the generalized eigenvalue problem. We compare our findings with predictions from the Nambu-Goto string model to assess potential tensions indicative of novel phenomena such as the existence of axion-like state along the flux-tube world-sheet. Notably, we provide undoubted evidence of the existence of a massive axion-like particle with the same mass as the corresponding axion extracted within the context of closed flux-tube. This strengthens the conjecture that the axion is a property of the world-sheet of the QCD string.}
\end{abstract}

\maketitle


\section{Introduction}

{\color{black} In the realm of confining gauge theories like $SU(N_c)$ gluodynamics with $N_c$ colours, the energy of a gauge field generated by colour charges is compressed into flux-tubes, forming what are known as confining strings. Unravelling the dynamics of the worldsheet associated with these confining strings stands as a pivotal endeavour in comprehending colour confinement. Despite considerable theoretical and lattice-based efforts, this inquiry remains unresolved. Nevertheless, progress has been made over the last decade, and concrete initiatives like the Simons Collaboration on Confinement and QCD Strings \cite{simons_confinement} are investing significant effort into unravelling this behaviour.}

{\color{black} Significant progress has been made over the past decade in understanding the effective string-theoretical description of the closed flux-tube~\cite{Athenodorou:2007du, Athenodorou:2010cs, Athenodorou:2011rx, Athenodorou:2024loq}. In particular, the dynamics of the worldsheet theory have been extracted from lattice data in a model-independent manner. This involves considering a long and smooth flux-tube, where the worldsheet dynamics are universal and governed by a low-energy theory of translational Goldstone modes (phonons). These dynamics are primarily described by the Nambu-Goto action ~\cite{Nambu:1978bd,Goto:1971ce,Zwiebach_2004}, supplemented by additional higher-dimensional operators~\cite{Aharony:2009gg}.

A standard method for calculating the excitation spectrum of torelons involves a perturbative expansion in powers of $1/(R\sqrt{\sigma})$, where $\sigma$ denotes the string tension~\cite{Luscher:2004ib, Aharony:2009gg} and $R$ the length of the string. This approach is effective for sufficiently long flux-tubes, as it neglects energy contributions from phonon scattering processes. Consequently, it is expected to work well for string states with no phonon content or for those containing a single phonon.

Recent advancements, however, propose that a Thermodynamic Bethe Ansatz (TBA) analysis, expanded in both $1/(R\sqrt{\sigma})$ and the softness of phonons (i.e., $p/\sqrt{\sigma}$), provides a more robust framework~\cite{Dubovsky:2013gi}. This approach offers an effective string theory capable of capturing the energy spectrum of the flux-tube across its entire length, ensuring greater accuracy even in regimes where traditional methods fall short.

A thorough TBA (Thermodynamic Bethe Ansatz) analysis of lattice data on closed flux-tubes has unveiled a fascinatingly rich structure in worldsheet dynamics. This includes the discovery of multiple string-like states and a significant massive excitation commonly referred to as the "worldsheet axion"~\cite{Dubovsky:2013gi, Athenodorou:2021vkw}. These insights highlight a complex interplay between the Goldstone modes and additional degrees of freedom present on the closed string worldsheet. This intricate dynamic has far-reaching implications, influencing both the formulation of low-energy effective field theories and our understanding of high-energy behaviour in such systems.

The axion appears to emerge naturally as a feature of the worldsheet action of the bosonic string. This raises the expectation that a similar property should manifest in the spectrum of the open flux-tube. Supporting evidence for this can be found in earlier work \cite{Morningstar:1996ze} and more recent studies \cite{Sharifian:2023idc}, where signals of such an excitation seem to be present. Specifically, states with masses corresponding to the absolute ground state in the $\Sigma_g^+$ representation are observed, coupled with an additional constant mass term.

This observation prompts a critical question: are these states intrinsic to the dynamics of the worldsheet, or do they result from couplings between the ground state and glueballs? Resolving this question requires computations in the large-$N_c$ limit~\cite{Lucini:2012gg,Manohar:1998xv,tHooft:1973alw}. Within this framework, phenomena like these are expected to be suppressed as non-planar contributions in the traditional large-$N_c$ simplification of connected correlation functions. Such an analysis would provide clearer insights into whether these excitations are a fundamental feature of the worldsheet dynamics or arise from interactions with external degrees of freedom.

Building upon the work reported in \cite{Sharifian:2023idc}, this study represents the first-ever consistent lattice calculation of the 3+1 dimensional open flux-tube spectrum in the large-$N_c$ limit. This marks a significant advancement by addressing potential caveats that could lead to an underestimation of the open-string spectrum.

Several improvements were implemented to ensure the reliability and precision of the results. For instance, anisotropic lattices were employed to improve the identification of mass plateaus, providing a more robust extraction of energy levels. Additionally, the fluctuations of the topological charge were carefully controlled to minimize systematic uncertainties by ensuring that topological freezing~\cite{Dromard:2014ela,Athenodorou:2021qvs,Bonanno:2022yjr} does not occur. A large basis of operators \cite{Capitani:2019,Muller:2019joq}was used to solve the Generalized Eigenvalue Problem (GEVP)~\cite{Blossier:2009kd}, ensuring a comprehensive exploration of the spectrum. Furthermore, advanced techniques such as smearing and multi-hit methods were utilized to enhance the signal-to-noise ratio, allowing for more accurate measurements of energy states.

These methodological refinements collectively establish a solid foundation for the accurate determination of the open flux-tube spectrum, providing new insights into the dynamics of confining strings in the large-$N_c$ limit.

This article is organized as follows: First, we discuss the expected theoretical predictions focusing on the Nambu-Goto string as well as Lorentz invariant Effective String Theories, followed by an explanation of the relevant quantum numbers describing the open flux-tube. Next, we provide a brief discussion on the lattice calculation, explaining how we obtain the flux-tube masses. Then we investigate the topological charge properties of the simulations demonstrating that our configuration ensembles are ergodic. Finally, we present our results for the spectrum of the open flux-tube and conclude.}

\section{Strings and the axionic string ansatz}

\begin{figure}[t!]
	\scalebox{1}{\includegraphics[width=9cm]{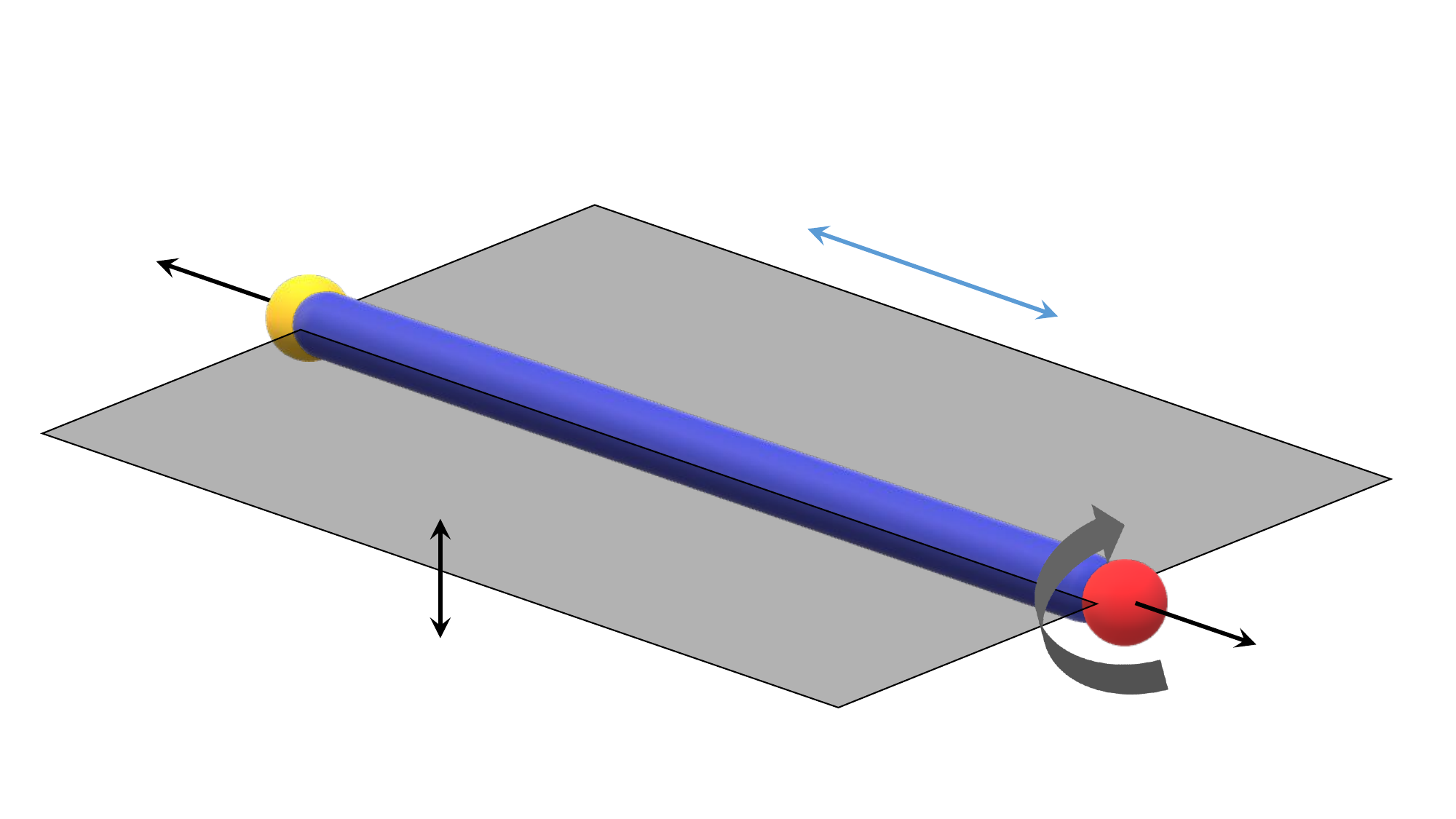}\put(-40,22){\Large $z$} \put(-200,15){\Large $\mathcal{P}_x = \pm$} \put(-50, 45){\Large $q$}  \put(-210, 100){\Large $\overline{q}$}\put(-90,100){\Large $\mathcal{C}o\mathcal{P} \equiv g/u$}\put(-70,10){ $\Lambda \to \Sigma,\Pi,\Delta,\Phi$}}
	\caption{Symmetries describing irreducibly the energy states of the flux-tube.\label{fig:flux_tube_symmetries}}
\end{figure}
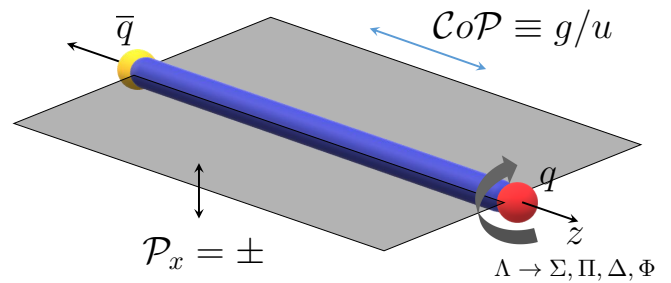

\subsection{The Nambu-Goto string}
\label{sec:Nambu_Goto}

{\color{black} A long flux-tube can be effectively described as a low-energy bosonic string. Its dynamics are governed by the translational Goldstone bosons associated with the breaking of \(ISO(1,D-1)\), the Poincaré symmetry in \(D\)-dimensional spacetime. This symmetry is spontaneously broken to the subgroup \(ISO(1,1) \times O(D-2)\), corresponding to longitudinal and transverse directions, respectively. As a result, the physics of the flux-tube is most naturally expressed using an action in terms of \(D\) worldsheet scalar fields, \(X^\mu\), which describe the embedding of the string's two-dimensional worldsheet into the \(D\)-dimensional target spacetime.

The leading-order action, which characterizes the simplest bosonic string model, is the Nambu-Goto action, given by:  
\begin{align}
S=-\sigma \int \mathrm{d}^2x \sqrt{ -\det h_{\alpha\beta}}\, ,
\end{align}
where \(\sigma\) is the string tension, and \(h_{\alpha\beta}\) is the induced metric on the string worldsheet, also referred to as the first fundamental form. The induced metric is defined as  
\begin{align}
h_{\alpha\beta} = \partial_\alpha X^\mu \partial_\beta X_\mu \,,
\end{align}
where \(\partial_\alpha X^\mu\) denotes the derivatives of the worldsheet embedding functions.

Quantizing the Polyakov action
\cite{Deser:1976rb,Brink:1976sc,Polyakov:1981rd,Polyakov:1987ez}, 
which is classically equivalent to the Nambu-Goto action, yields the energy spectrum of an open relativistic string of length \(R\) with fixed endpoints. The resulting spectrum is described by the Arvis potential \cite{ARVIS1983106}:  
\begin{align}
E^{\rm NG}_N(R) = \sqrt{\sigma^2 R^2 + 2\pi\sigma \left(N - \frac{D-2}{24}\right)} \,,
\label{eq:Nambu_Goto}
\end{align}
where \(N\) defined in Eq.~\ref{eq:occupation_number}, not to be confused with the number of colours $N_c$, is the principal quantum number associated with the string's vibrational modes, and \(D\) is the dimension of spacetime. Since $E^{\rm NG}_N(R)$ is commonly referred to as the Arvis potential, we will adopt the notation $V(R) \equiv E^{\rm NG}_N(R)$ throughout this manuscript.

This expression captures the relativistic corrections to the string's energy and demonstrates how the string tension and quantum fluctuations combine to determine its behaviour at both classical and quantum levels. The factor \({(D-2)}/{24}\) accounts for the contributions of the zero point energies of the \(D-2\) transverse Goldstone bosons and reflects the string's quantum nature in \(D\)-dimensional spacetime. Of course Weyl anomaly suggests that from a first sight Nambu-Goto energy prediction does not preserve Lorentz invariance, thus, hindering a concrete interpretation for the confining flux-tube. The challenge of formulating a Lorentz-invariant description of the confining flux-tube has engaged physicists for over four decades.}

\begin{table}[t!]
\begin{ruledtabular}
	\begin{tabular}{c|c|cc}
		Excitation &
		Symmetry&
	State&\\
		\colrule  \hline
		$N=0$& $\Sigma_g^+$& $|0\rangle$& \\
		\colrule
		$N=1$& $\Pi_u$& $a^\dagger_{1^+}|0\rangle$& $a^\dagger_{1^-}|0\rangle$ \\
		\colrule
		$N=2$ & ${\Sigma_g^{+}}'$& $a^\dagger_{1^+} a^\dagger_{1^-}|0\rangle$&  \\
		& $\Pi_g$& $a^\dagger_{2^+}|0\rangle$& $a^\dagger_{2^-}|0\rangle$ \\
		& $\Delta_g$& $(a^\dagger_{1^+})^2|0\rangle$& $(a^\dagger_{1^-})^2|0\rangle$ \\
		\colrule
		$N=3$& $\Sigma_u^{+}$& $(a^\dagger_{1^+}a^\dagger_{2^-}+a^\dagger_{1^-}a^\dagger_{2^+})|0\rangle$&  \\
	& $\Sigma_u^-$& $(a^\dagger_{1^+}a^\dagger_{2^-}-a^\dagger_{1^-}a^\dagger_{2^+})|0\rangle$&  \\
		& $\Pi_u'$& $a^\dagger_{3^+}|0\rangle$& $a^\dagger_{3^-}|0\rangle$ \\
		& $\Pi_u''$& $(a^\dagger_{1^+})^2 a^\dagger_{1^-}|0\rangle$&  $a^\dagger_{1^+} (a^\dagger_{1^-})^2|0\rangle$ \\
			& $\Delta_u$& $a^\dagger_{1^+} a^\dagger_{2^+}|0\rangle$&  $a^\dagger_{1^-} a^\dagger_{2^-}|0\rangle$\\
			&$\phi_u$& $(a_{1^+}^\dagger)^3|0\rangle$&  $(a_{1^-}^\dagger)^3|0\rangle$\\
		\colrule
		$N=4$ & ${\Sigma_g^{+}}''$& $a^\dagger_{2^+} a^\dagger_{2^-}|0\rangle$&  \\
	 & ${\Sigma_g^{+}}'''$& $(a^\dagger_{1^+})^2 (a^\dagger_{1^-})^2|0\rangle$&  \\
	  & ${\Sigma_g^{+(\mathrm{iv})}}$& $(a^\dagger_{1^+}a^\dagger_{3^-}+a^\dagger_{1^-}a^\dagger_{3^+})|0\rangle$&  \\
	   & ${\Sigma_g^{-}}$& $(a^\dagger_{1^+}a^\dagger_{3^-}-a^\dagger_{1^-}a^\dagger_{3^+})|0\rangle$  \\
	   &$\Pi_g'$&$a^\dagger_{4^+}|0\rangle$& $a^\dagger_{4^-} |0\rangle$\\
	   &$\Pi_g^{''}$&$(a^\dagger_{1^+})^2 a^{\dagger}_{2^-}|0\rangle$& $(a^\dagger_{1^-})^2 a^{\dagger}_{2^+}|0\rangle$\\
	   &$\Pi_g^{'''}$&$a^\dagger_{1^+} a^\dagger_{1^-} a^{\dagger}_{2^+}|0\rangle$& $a^\dagger_{1^+} a^\dagger_{1^-} a^{\dagger}_{2^-}|0\rangle$\\
	   &$\Delta_g'$&$a^\dagger_{1^+}a^\dagger_{3^+}|0\rangle$&$a^\dagger_{1^-}a^\dagger_{3^-}|0\rangle$\\
	   &$\Delta_g^{''}$&$(a^\dagger_{2^+})^2|0\rangle$&$(a^\dagger_{2^-})^2|0\rangle$\\
	   &$\Delta_g^{'''}$&$(a^\dagger_{1^+})^3a^\dagger_{1^-}|0\rangle$&$a^\dagger_{1^+}(a^\dagger_{1^-})^3|0\rangle$\\
	   &$\Phi_g$&$(a^\dagger_{1^+})^2a^\dagger_{2^+}|0\rangle$&$(a^\dagger_{1^-})^2a^\dagger_{2^-}|0\rangle$\\
	   &$\Gamma_g$&$(a^\dagger_{1^+})^4|0\rangle$&$(a^\dagger_{1^-})^4|0\rangle$\\
	\end{tabular}
\end{ruledtabular}
\caption{\color{black}Low-lying string states for an open string with fixed ends expressed in terms of right-handed (+) and left-handed (–) circularly polarized phonons, following the notation in Ref.~\cite{Juge:2003ge}. We note that the number of prime symbols appearing as superscripts in the notation of the irreducible representations indicates the excitation level.
\label{table:quantum_numbers}}
\end{table}

\subsection{Lorentz invariant Effective String Theories}
\label{sec:lorentz_EST}
{\color{black} Systematic methods for studying Lorentz-invariant Effective String Theory (EST), which describes the confining string, were first developed by Lüscher, Symanzik, and Weisz in \cite{Luscher:1980ac} using the static gauge, and by Polchinski and Strominger in \cite{Polchinski:1991ax} using the conformal gauge. These approaches yield predictions for the energy spectrum of states as an expansion in \( 1/R\sqrt{\sigma} \). Specifically, terms of order \( O(1/R^p) \) arise from \((p+1)\)-derivative contributions to the EST action, whose coefficients correspond to Low Energy Coefficients (LECs) that are initially unconstrained. However, it was later discovered that these LECs are subject to stringent constraints due to the non-linear realization of Lorentz symmetry \cite{Luscher:2004ib,Meyer:2006qx,Aharony:2009gg}. As a result, certain terms in the \( 1/R \) expansion can be predicted without any free parameters.

Different formulations of Effective String Theory (EST) are distinguished by the choice of gauge fixing applied to the embedding coordinates on the world-sheet. Two primary approaches exist: the {\it static gauge}~\cite{Luscher:1980ac,Luscher:2004ib,Aharony:2009gg} and the {\it conformal gauge}~\cite{Polchinski:1991ax,Drummond:2004yp,HariDass:2009ub} as mentioned in the previous paragraph, both of which ultimately yield the same physical predictions.  

At the foundation of EST lies the leading-order area term, which governs the large-string behaviour and gives rise to a linearly rising potential, expressed as \(E \simeq \sigma R\). The next contribution comes from the Gaussian action, responsible for the well-known Lüscher term, proportional to \(1/R\), whose coefficient is universal and depends only on the space-time dimension \(D\):
\begin{eqnarray}
    S_{EST} = \sigma R T +\frac{\sigma}{2}\int \mathrm{d}^2 x \partial_\alpha X_i . \partial^\alpha X^i + \ldots
\end{eqnarray}
Higher-order corrections emerge from terms with increasing numbers of derivatives i.e:
\begin{eqnarray}
    +\frac{\sigma}{2}\int \mathrm{d}^2 x \left(  \frac{1}{8} ( \partial_\alpha X_i . \partial^\alpha X^i)^2  - \frac{1}{4} ( \partial_\alpha X_i . \partial^\beta X^i)^2 \right).
\end{eqnarray}
The above four-derivative term introduces a correction proportional to \(1/R^3\), again featuring a universal coefficient that depends on \(D\). Extending this framework to six-derivative terms~\cite{Aharony:2010cx,Aharony:2010db}, one finds that in \(D=3\), these terms generate a universal correction proportional to \(1/R^5\). However, for general excited states in \(D=4\), the coefficient of the \(O(1/R^5)\) term is not universal, though notably, the correction remains universal for the ground state. 

Unlike the closed-flux-tube case, where one must account for derivative terms arising from the dynamics of the oscillating bosonic string, the open flux-tube case requires considering boundary terms as well~\cite{Aharony:2010cx,Aharony:2010db}.

In summary, the energy spectrum for a long string follows the expansion:
{\small
  \begin{eqnarray}
    E_N(R)  & = & \sqrt{\sigma} (R  \sqrt{\sigma}) +  \frac{\pi \sqrt{\sigma}}{(R  \sqrt{\sigma})} \left( N - \frac{D-2}{24}  \right) \nonumber \\
            & - & \frac{ \pi^2 \sqrt{\sigma}}{2(R  \sqrt{\sigma})^3} \left( N - \frac{D-2}{24}  \right)^2 \nonumber \\
            & + & \frac{\bar{b}_2 \pi^3 \sqrt{\sigma}}{(R  \sqrt{\sigma})^4} \left( B_N - \frac{D-2}{60}  \right) \nonumber \\
            & + & \frac{\pi^3 \sqrt{\sigma}}{16(R  \sqrt{\sigma})^5} \left( N - \frac{D-2}{24}  \right)^3 \nonumber \\
            & + & \frac{\pi^3 \sqrt{\sigma} (D-26)}{48 (R  \sqrt{\sigma})^5}C_N+
            \left(  \frac{1}{(R  \sqrt{\sigma})^6}   \right).
\end{eqnarray}}
The leading term, along with the terms proportional to \( {1}/{R\sqrt{\sigma}} \), \( {1}/{(R\sqrt{\sigma})^3} \), and the first term proportional to \( {1}/{(R\sqrt{\sigma})^5} \), matches the expansion of Equation~\ref{eq:Nambu_Goto} up to order \( O\left({1}/{(R\sqrt{\sigma})^7}\right) \), while the terms with coefficients \( \bar{b}_2 \) and \( C_N \) arise from boundary effects.
 Therefore, the Nambu-Goto equation can be interpreted as a particular resummation of tree-level terms in the \( 1/(R\sqrt{\sigma}) \) expansion. Therefore, the energy of the open flux-tube takes the following expression:
\begin{eqnarray}
\label{eq:Nambu_goto_b2}
    E_N(R)  & = &   E^{\rm NG}_{N}(R) +  \frac{\bar{b}_2 \pi^3 \sqrt{\sigma}}{(R  \sqrt{\sigma})^4} \left( B_N - \frac{D-2}{60}  \right) \nonumber \\ 
    & + &  \frac{\pi^3 \sqrt{\sigma} (D-26)}{48 (R  \sqrt{\sigma})^5}C_N  +  \left(  \frac{1}{(R  \sqrt{\sigma})^6}  \right).
\end{eqnarray}
In the above expression, \(\bar{b}_2\) is a universal coefficient that does not depend on the excitation number \(N\), whereas \(B_N\) and \(C_N\) vary with \(N\). In this work, we will attempt to extract $\bar{b}_2$ the subleading \(1/R^4\) term, for which \(B_0 = 0\) and \(B_1 = 4\).}

\subsection{The string Quantum Numbers}
\label{sec:quantum_numbers}

{\color{black}
\begin{figure}[t!]
   \begin{minipage}{\columnwidth}
	\includegraphics[scale=0.85]{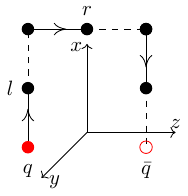}
\subcaption{Staples used for the construction of spatial Wilson lines in order to project onto flux-tube states described by $\Sigma_g^+$,  $\Pi_u$, and $\Delta_g$.\label{fig:op_oi}}
    \end{minipage}\vfil
   \begin{minipage}{\columnwidth}
	\includegraphics[scale=0.85]{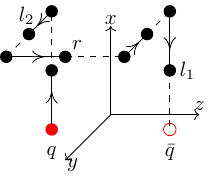}
\subcaption{Staples used for the construction of spatial Wilson lines in order to project onto flux-tube states described by $\Sigma_g^-$.}
   \end{minipage}\vfil
   \begin{minipage}{\columnwidth}
     \centering
	\includegraphics[scale=0.85]{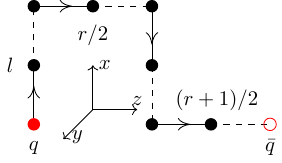}
\subcaption{Staples used for the construction of spatial Wilson lines in order to project onto flux-tube states described by $\Sigma_u^+$, ${\Pi_g}$, and  ${\Delta_u}$.}
    \end{minipage}\vfil
   \begin{minipage}{\columnwidth}
     \centering
	\includegraphics[scale=0.85]{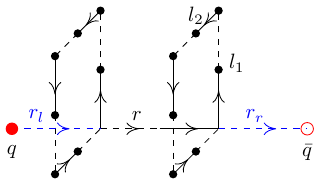}
\subcaption{Staples used for the construction of spatial Wilson lines in order to project onto flux-tube states described by $\Sigma_u^-$. \label{fig:op_loop}}   
\end{minipage}

\caption{Operators are built using the staples above based on the prescription described in Ref.~\cite{Sharifian:2023idc}. Once assembled to irreducibly project onto a particular symmetry, the operators are used on the fly during the simulation to compute the entries of the Wilson correlation matrix (see Fig.~\ref{fig:wilson_loop}). \label{fig:sub-operators}}
\end{figure}

 To achieve a comprehensive understanding of the spectrum of the open flux-tube, it is essential to extract the energy levels of all states that can be constructed from all possible combinations of phonon operators acting on the vacuum. In practice, this requires projecting onto all states that transform according to the relevant quantum numbers associated with the symmetries of the system.

The energy states of the open flux-tube can be classified irreducibly using three key quantum numbers. These symmetries are illustrated schematically in Figure~\eqref{fig:flux_tube_symmetries} and described in detail below:
\begin{enumerate}
\item {\bf Angular Momentum Projection (\( \Lambda \))}:  
   The first symmetry corresponds to the projection of angular momentum \( J \) onto the charge axis, represented by \( J \cdot \hat{R} \), where \( \hat{R} \) is the unit vector along the charge axis. This projection is denoted by \( \Lambda \). It is conventional to use Greek letters to label these states, with \( \Sigma, \Pi, \Delta, \Phi, \ldots \) corresponding to \( \Lambda = 0, 1, 2, 3, \ldots \), respectively. This notation is inspired by the irreducible representations of the dihedral group commonly used in molecular symmetry classifications~\cite{atkins2017physical}.

\item {\bf Charge Conjugation and Parity (\( \eta_{CP} \))}:  
   The second symmetry involves the combination of charge conjugation (\( \mathcal{C} \)) and spatial inversion (\( \mathcal{P} \)) about the midpoint between the quark and the antiquark. This combined operation, denoted as \( \mathcal{C}o\mathcal{P} \), has eigenvalues \( \eta_{CP} \), which are labeled as \( g \) or \( u \), corresponding to \( +1 \) (even) or \( -1 \) (odd) eigenvalues, respectively.
\item {\bf Reflection Symmetry for \( \Sigma \) States (\( \epsilon \))}:  
   For \( \Sigma \) states (\( \Lambda = 0 \)), an additional quantum number \( \epsilon \) is defined, representing the eigenvalue of the reflection operator with respect to any plane containing the charge axis. These eigenvalues are labeled as \( + \) for even and \( - \) for odd reflection symmetry. It is important to note that for states with \( \Lambda \geq 1 \), the reflection symmetry does not affect the energy of the gluonic excitations. Instead, such reflections merely change the handedness of the state, as indicated by the sign of \( J \cdot \hat{R} \).
\end{enumerate}
Based on these symmetries, the states of the open flux-tube are categorized and denoted as \( \Sigma_g^+ \), \( \Sigma_g^- \), \( \Sigma_u^+ \), \( \Sigma_u^- \), \( \Pi_g \), \( \Pi_u \), \( \Delta_g \), \( \Delta_u \), and so on. This systematic classification provides a detailed framework for analysing the spectrum of open flux-tubes and the associated gluonic excitations.}

\subsection{Quantum Numbers and phonons}
\label{sec:Quantum_Numbers}

{\color{black} As expected, the phononic content along the fluctuating string results in stationary states, which can be expressed in terms of normal modes using the \( D-2 \) transverse displacement fields \( X^i \). These modes carry relative momenta \( m\omega \) for positive integer \( m \), where \( \omega = \pi/R \). 

We define right-(\(+\)) and left-(\(-\)) circularly polarized ladder operators \( a_{m_\pm}^\dagger \), which, for simplicity, we refer to as "phonons." The string states can then be constructed as linear combinations of eigenmodes, expressed as:

\begin{align}
\prod_{m=1}^{\infty}	\left(( a_{m^+})^{n_{m_+}} 	( a_{m^-}^\dagger)^{n_{m_-}}\right)|0\rangle, 
\end{align}
where \( |0\rangle \) denotes the ground state of the string, and \( n_{m^+} \) and \( n_{m^-} \) are occupation numbers taking values \( 0, 1, 2, \dots \). The eigenvalues of the principal quantum number \( N \), the angular momentum quantum number \( \Lambda \), and parity \( \eta_{CP} \) can be determined using the following expressions:
\begin{align}
\label{eq:occupation_number}
&N=\sum_{m=1}^{\infty} m (n_{m^+}+n_{m^-}),\\
&\Lambda=\big|\sum_{m=1}^\infty \left(n_{m^+}-n_{m^-}\right)\big|, \\
&\eta_{CP}=(-1)^N.
\end{align}
In Table \eqref{table:quantum_numbers}, we classify the flux-tube states based on their energy and the symmetries governing the open flux-tube. For each combination of excitation level and molecular symmetry, we provide the associated string states. As evident from the Table, multifold degeneracies emerge as energy levels increase.

\newpage

\begin{table*}[t!]
	\centering
\begin{tabular}{c|cccccccccc}
Ensemble&$N_c$&$\beta$&$\xi_0$&$\xi_r$&lattice&$a_s\sqrt{\sigma}$&$a_t\sqrt{\sigma}$&$\langle Plaquette \rangle$&$\chi^{1/4}/\sqrt{\sigma}$&$\#config$
\\
\hline\hline
$W_{3, 2}$&$3$&$5.9$&$2$&$2.1735$&$24^3\times48 $&$0.3049(17)$&$0.1403(8)$&$0.595665(6)$&$0.4239(4)$&$1000$
\\
$W_{4, 2}$&$4$&$10.7$&$2$&$2.1825$&$24^3\times48 $&$0.3493(9)$&$0.1601(4)$&$0.571288(5)$&$0.4232(6)$&$1000$
\\
$W_{5, 2}$&$5$&$17.2$&$2$&$2.1830$&$24^3\times48 $&$0.3006(5)$&$0.1377(2)$&$0.575065(4)$&$0.4021(5)$&$1000$
\\
$W_{6, 2}$&$6$&$24.9$&$2$&$2.1850$&$24^3\times48 $&$0.3107(15)$&$0.1422(7)$&$0.568885(3)$&$0.3943(6)$&$1000$
\\
$W_{3, 4}$&$3$&$5.7$&$4$&$4.5385$&$24^3\times96 $&$0.4024(16)$&$0.0887(4)$&$0.617109(6)$&$0.4395(1)$&$1000$
\\
$W_{4, 4}$&$4$&$10.4$&$4$&$4.5663$&$24^3\times96 $&$0.4445(3)$&$0.0973(1)$&$0.599735(5)$&$0.4477(7)$&$1000$
\\
$W_{5, 4}$&$5$&$16.5$&$4$&$4.5766$&$24^3\times96 $&$0.4484(1)$&$0.0980(0)$&$0.593900(4)$&$0.4415(10)$&$1000$
\\
$W_{6, 4}$&$6$&$24.0$&$4$&$4.5811$&$24^3\times96 $&$0.4333(2)$&$0.0946(0)$&$0.592902(4)$&$0.4434(10)$&$1000$
\\
\hline\hline
\end{tabular}

\caption{Details of the ensembles generated using the anisotropic Wilson action. $\xi_r$ is the renormalized anisotropy factor, computed using the method described in Ref.~\cite{GarciaPerez:1996ft}. All ensembles have undergone 100 times multihit smearing in the temporal direction and APE smearing ($\alpha = 0.3$, $n_s = 20$) in the spatial directions.} \label{tab:ensemble}
\end{table*}}

\subsection{The Axionic String Ansatz (ASA)}
\label{subsec:ASA}

{\color{black} In the case of closed flux-tubes, the universal predictions of the Thermodynamic Bethe Ansatz (TBA) have been shown to accurately describe a large sector of confining string excitations. This has been confirmed with high precision in both $D=4$ and $D=3$ dimensions \cite{Athenodorou:2010cs,Athenodorou:2011rx,Athenodorou:2018sab,Athenodorou:2022pmz}. However, at shorter string lengths $R$, most states exhibit increasing deviations from these predictions, indicating the influence of non-universal corrections. TBA analysis has allowed for the determination of the first non-trivial Wilson coefficient for $D=3$ confining strings~\cite{Dubovsky:2014fma,Chen:2018keo}. Interestingly, some $D=4$ states including the ground state with quantum numbers $0^{--}$ show significant deviations from universal TBA predictions even at relatively large $R$, suggesting the presence of an additional massive excitation on the string worldsheet, commonly referred to as the "string axion" \cite{Dubovsky:2013gi,Dubovsky:2014fma}.

This suggests the existence of an additional term in the worldsheet action that governs the leading-order interactions between the worldsheet axion and the transverse Goldstone modes. The corresponding action is given by:
\begin{align}
    S_\phi= \int d^2 \sigma \sqrt{-h} &\bigg( -\frac{1}{2} (\partial \phi)^2 - \frac{1}{2} m^2 \phi^2 \nonumber\\&+ \frac{Q_{\phi}}{4} h^{\alpha \beta} \epsilon_{\mu \nu \lambda \rho} \partial_\alpha t^{\mu \nu} \partial_\beta t^{\lambda \rho} \phi \bigg) ,
    \label{axion_interaction}
\end{align}
where \(\phi\) represents the pseudoscalar worldsheet axion, and \(t^{\mu \nu} = \frac{\epsilon^{\alpha \beta}}{\sqrt{-h}} \partial_\alpha X^\mu \partial_\beta X^\nu\) is the extrinsic curvature of the worldsheet. The axion mass \(m\) can be determined from Monte Carlo simulations of the 4D \(SU(N_c)\) Yang-Mills confining flux-tube spectrum using TBA analysis, yielding values of \(m=1.82(2)\) for $SU(3)$, \(m=1.65(2)\) for $SU(5)$ and  \(m=1.65(2)\) for $SU(6)$ \cite{Athenodorou:2024loq}.

\begin{figure}[t!]
	\centering
	\includegraphics[scale=1.15]{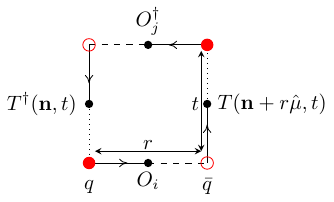}
	\caption{A closed loop corresponding to the entry $\mathcal{C}_{i,j}(r,t)$ of the Wilson correlation matrix.\label{fig:wilson_loop}}
\end{figure}}

\section{Lattice QCD methodology\label{sec:lat_qcd}}
In this section,  we outline the lattice QCD framework  used in our calculations. 

\subsection{Our lattice framework to study the flux-tube spectrum}
\label{sec:lattice_framework}

{\color{black}

\begin{figure*}[t!]
 \hspace{-3.0cm}
    \begin{minipage}{\columnwidth}
        \includegraphics[scale=0.8]{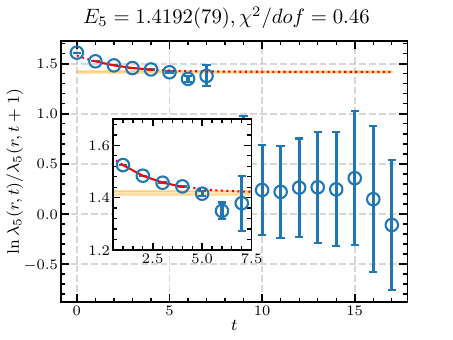}
    \end{minipage} \hspace{-3cm}
    \begin{minipage}{\columnwidth}
        \includegraphics[scale=0.8]{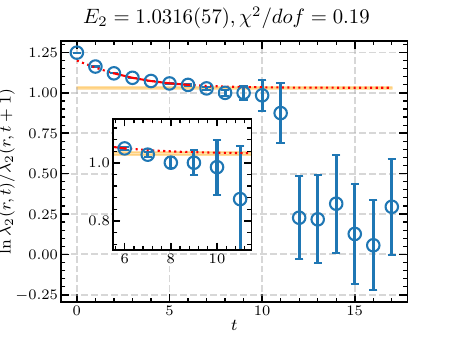}
    \end{minipage} \hspace{-3cm}
    \begin{minipage}{\columnwidth}
        \includegraphics[scale=0.77]{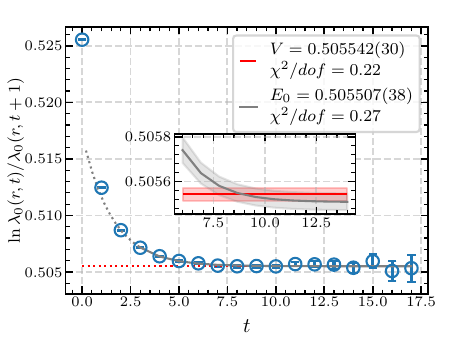}
    \end{minipage} \hspace{-3cm}
    \caption{
Effective mass determination using fits to Eq.~\ref{eq:multi-exp-ansatz}. The fitted curve is shown as a solid line, while the dotted line represents the extrapolation of the fit. The width of the shaded band indicates the uncertainty in the effective mass. The middle and left panels illustrate cases where the signal-to-noise ratio is large, and the plateau is not present. In contrast, the right panel shows a case where a plateau is clearly identified, and the result is consistent with a fit to Eq.~\ref{eq:multi-exp-ansatz}.
\label{fig:multi_exp_fit}
}
\end{figure*}

In this work, we focus on the large-distance behaviour of the excited spectrum, deliberately avoiding short-distance effects. To achieve this, we employ the Wilson action discretized on anisotropic lattices \cite{Morningstar:1996ze}, which is expressed as:
\begin{equation}
	S_\text{Wilson} = \beta\left(\frac{1}{\xi} \sum_{x,s>s'} W_{s,s'} + \xi \sum_{x,s} W_{s,t} \right), \label{eq:wilsonAniso}
\end{equation}
where $\beta = 2N_c/g^2$ is the inverse coupling constant and $\xi$ represents the bare anisotropy factor, defined as the ratio of the spatial to temporal lattice spacings $(a_s/a_t)$. Furthermore, with $s$, $s'$ we denote spatial links in different directions. Here, $W_{s,s'}$ represents the spatial plaquettes, and $W_{s,t}$ the spatial-temporal plaquettes. 
By using an anisotropic action, we achieve finer lattice spacing in the temporal direction while maintaining a coarser spacing in  spatial directions.  
This results in more data points for the effective mass plot along the time direction and enables the study of larger intercharge distances. For investigations at short distances see for instance \cite{Juge:2002br,Karbstein:2014bsa,Schlosser:2021wnr,Schlosser:2025tca}.

In this work, we use the publicly available code introduced in Ref.~\cite{Cardoso:2011xu} to generate $SU(N_c)$ lattice QCD pure gauge configurations. This code is developed using CUDA to run on GPUs. The parameters of each configuration ensemble produced are listed in Table~\ref{tab:ensemble}. The selected lattice parameters ensure that the measurements are free from finite volume effects~\cite{Sharifian:2023}. To ensure ergodicity, we examine the topological charge and its evolution. The topological aspects of the calculation are presented in Section~\ref{sec:topology}. }

{\color{black} As discussed in the previous paragraph, maintaining a coarse lattice spacing requires increasing the resolution along the temporal extent. To achieve this, we consider two values of the spatial lattice spacing for each choice of the number of colors: \( a_s \sqrt{\sigma} \sim 0.4 \) with a bare anisotropy of \( \xi = 4 \), and \( a_s \sqrt{\sigma} \sim 0.3 \) with a bare anisotropy of \( \xi = 2 \). The value of \( \beta \) is expected to scale approximately as \( \sim N^2_c \), and its specific values have been chosen based on the results of Ref.~\cite{Lucini:2004my}. 
{\color{black}
Since a significant tuning effort would be necessary to have exactly identical lattice spacings for all different $N_c$, the matching between our different spatial lattice spacings remains approximate.

To compute the lattice spacing, we fit the Cornell potential to the ground state \(\Sigma_g^+\):
\[
V(N_s) = V_0 + \frac{B}{N_s} + C N_s,
\]
where \( N_s \) is the spatial separation. We fit for large values of  \( N_s \) for which higher corrections in \( 1/ N_s \) are negligible. Since we present our data in string tension units, the lattice spacings are given by:
\[
a_s \sqrt{\sigma} = \sqrt{C \, \xi_r}, \qquad a_t \sqrt{\sigma} = \sqrt{\frac{C}{\xi_r}},
\]
where \( \xi_r \) is the renormalized anisotropy.

The renormalized anisotropy factor $\xi_r$ was perturbatively computed at one-loop order by Ref.~\cite{GarciaPerez:1996ft}, extending Ref.~\cite{Karsch:1982ve} and using the finite volume effective action for $SU(N_c)$ gauge theories in the background of a zero-momentum gauge field. An alternative approach consists of computing non-perturbatively the ratio of the spatial-temporal and spatial-spatial Wilson loops. This was used in Ref. ~\cite{Klassen:1998ua} at short distances for the $SU(3)$ case and in Ref.~\cite{Amado:2013rja} at longer distances (with the string tensions) for the $U(1)$ gauge theory. Ref.~\cite{Drummond:2002yg}  verified the agreement of the two approaches for $SU(3)$ within discrepancies of less than 3\%. Since we are interested in $SU(N_c)$, we utilize the results of  Ref.~\cite{GarciaPerez:1996ft}, leaving for future studies the direct computation on the lattice of the $SU(N_c)$ renormalized anisotropies.

Specifically, the expression for the  normalized anisotropy factor $\xi_r$ is given by
\begin{equation}
    \frac{\xi_r(N_c,\xi,\beta)}{\xi} = 1 + \eta_1(N_c,\xi) / \beta  + \cdots \ , 
\label{eq:renorm}
\end{equation}
where,
\be
\eta_1(N_c,\xi) = \frac{ 9 -{N_c}^2}{ 5} \eta_1(2,\xi) + \frac{ {N_c}^2  - 4 }{5} \eta_1(3,\xi) \ .
\label{eq:su2su3}
\ee
The function $\eta_1(N_c,\xi)$ is determined by the cases of $SU(2)$ and $SU(3)$ and these are provided with a very high precision in Table 1 of Ref.~\cite{GarciaPerez:1996ft}, in Table.~\ref{Tab:eta1} we show the values related to our work. 
\begin{table}
\centering
\begin{ruledtabular}
\begin{tabular}{cccc}
Group & $N$ & $\xi$ & $\eta_1(\xi, N)$ Wilson \\
\hline\hline
$SU(2)$ & 2 & 2 & 0.18006435 \\
$SU(2)$ & 2 & 4 & 0.26388176 \\
$SU(3)$ & 3 & 2 & 0.51182234 \\
$SU(3)$ & 3 & 4 & 0.76739428 \\

\end{tabular}
\end{ruledtabular}
\caption{Values of $\eta_1(N,\xi)$ from Ref.~\cite{GarciaPerez:1996ft}, used as input in Eq.~(\ref{eq:su2su3}) to compute the renormalized anisotropy $\xi_r$ in this work. \label{Tab:eta1}}
\end{table}
Interestingly, in the case of the Wilson action, these functions are  approximately linear in $1/ \xi$. We consider the cases of $\xi=2$ and $\xi=4$ and the respective one-loop renormalized anisotropies are presented in Table~\ref{tab:ensemble}.
}

The operators are constructed using smeared links. Specifically, we apply APE~\cite{Albanese:1987ds,Morningstar:2003gk} smearing to spatial links, while temporal links are smeared using the multihit technique~\cite{Huntley:1985ts,Bicudo:2018jbb}.

We use linear combinations of the staples shown in Fig. \ref{fig:sub-operators} consisting of rotations around the principal axis and reflections over the two parity planes to compute the Wilson correlation matrix for each irreducible representation of the flux-tube.

\subsection{Extracting the effective mass}

To compute the spectrum of the flux-tube, we first construct the Wilson correlation matrix, $\mathcal{C}(r, t)$. The entry $\mathcal{C}_{i,j}(r, t)$ of this matrix represents the expectation value of space-time closed loops (see Fig.~\ref{fig:wilson_loop}), where the spatial segments are replaced by operators $O_i$ and $O_j$ that share the same symmetry as the flux-tube of interest.

Next, we compute the generalized eigenvalues $\lambda_n$ \cite{Blossier:2009kd,Dudek:2009qf,Dudek:2010wm,Bicudo:2021qxj} of the Wilson correlation matrix by solving  

\begin{align}
	\mathcal{C}(r, t) \vec{\nu}_n = \lambda_n(r, t) \mathcal{C}(r, t_0) \vec{\nu}_n, 
\end{align}  

where we set $t_0 = 0$. This yields a set of time-dependent eigenvalues $\lambda_n(t)$ for each $r$, which we then arrange in ascending order, with $n$ indicating their rank in the ordering. The effective mass is then obtained as the plateau observed in

\begin{equation}
    E_n = \ln \frac{\lambda_n(r, t)}{\lambda_n(r, t+1)}.
\end{equation}

In practice, finding a clear plateau is not always possible because larger values of $r$ and $t$ introduce significant noise. As a result, the plateau may either not exist or be unreliable. By closely examining $\ln \frac{\lambda_n(r,t)}{\lambda_n(r,t+1)}$, we observe that the plot is smooth for small values of $t$, but contributions from excited states still persist.

For simplicity, let us consider how the Wilson loop is related to the energy of the states,   
\begin{equation}
    \langle W(r,t)\rangle=\sum_{n=0}^{\infty} A_n e^{-E_n(r) t},
\end{equation}
where we assume that $E_n$ are ordered in ascending values.  We aim to determine $E_n$. To eliminate a degree of freedom, it is common to consider  
\begin{equation}
\frac{W(r, t)}{W(r, t+1)}=e^{E_0} \frac{1+A_1 e^{-(E_1-E_0)t}+\ldots}{1+A_1 e^{-(E_1-E_0)(t+1)}+\ldots}.\label{eq:multi-exp-fit}
\end{equation}
Typically, the energy values are extracted from the plateau in the effective mass plot for large $t$, where the higher-order exponential terms become negligible. The effective energy plateau value is defined as: 
\begin{equation}
E_0(r,t)=\lim_{t\to \infty} \ln \frac{W(r,t)}{W(r,t+1)}.
\end{equation}
For cases where the plateau is absent or unreliable, we retain only the first or second exponential term in Eq.~\ref{eq:multi-exp-fit} and fit the ansatz  
\begin{equation}
    \ln\frac{W(r,t)}{W(r,t+1)}=E_0 +\ln \frac{1+A_1 e^{-\delta E t}}{1+A_1 e^{-\delta E(t+1)}}.\label{eq:multi-exp-ansatz}
\end{equation}
This approach requires at least four data points to fit the ansatz. In Fig. \ref{fig:multi_exp_fit}, we illustrate how this method works. In the last plot of Fig. \ref{fig:multi_exp_fit}, we compare this method with the effective mass plateau on an ensemble where such a comparison is feasible, finding consistent results between the two approaches. Our primary goal of course, is to identify the plateau; however, when no plateau is found, we extract the potential using Eq.~\ref{eq:multi-exp-ansatz}. This situation commonly occurs in the effective mass plots of more excited states, where no clear signal is visible.

\begin{figure*}[t!]
\centering
    \begin{minipage}{\columnwidth}
        \includegraphics[scale=0.8]{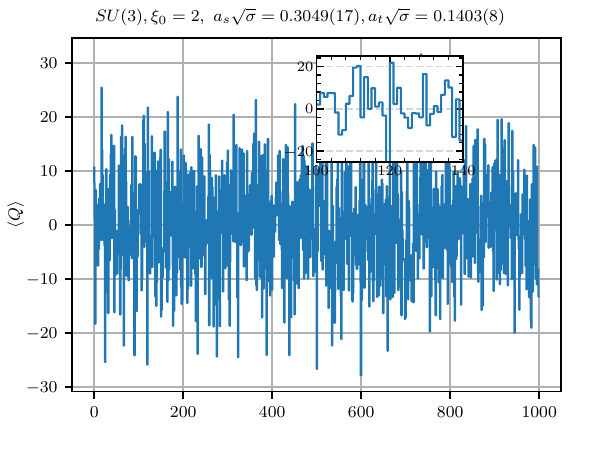}
    \end{minipage}\hfil
      \begin{minipage}{\columnwidth}
        \includegraphics[scale=0.8]{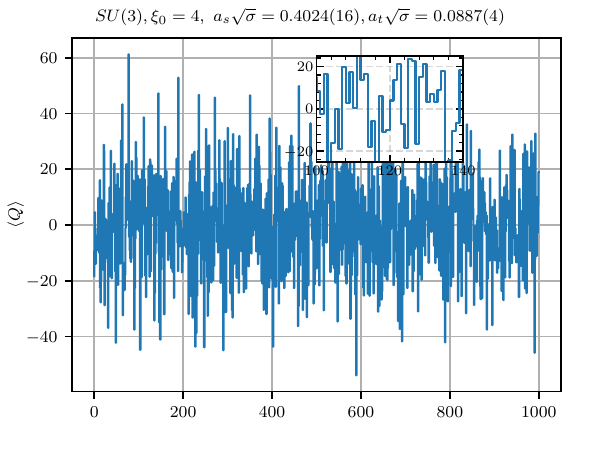}
    \end{minipage}\vfil
        \begin{minipage}{\columnwidth}
        \includegraphics[scale=0.8]{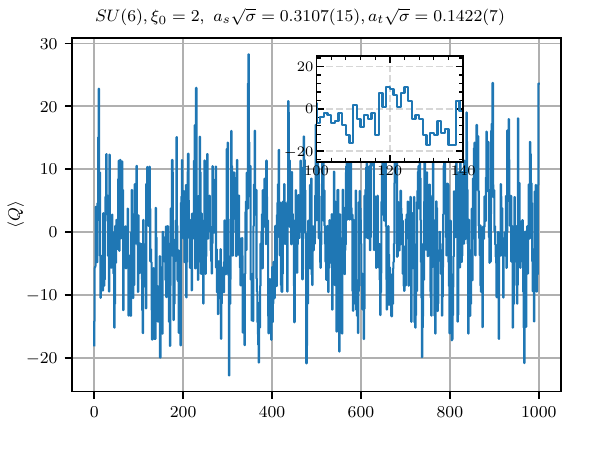}
    \end{minipage}\hfil
         \begin{minipage}{\columnwidth}
        \includegraphics[scale=0.8]{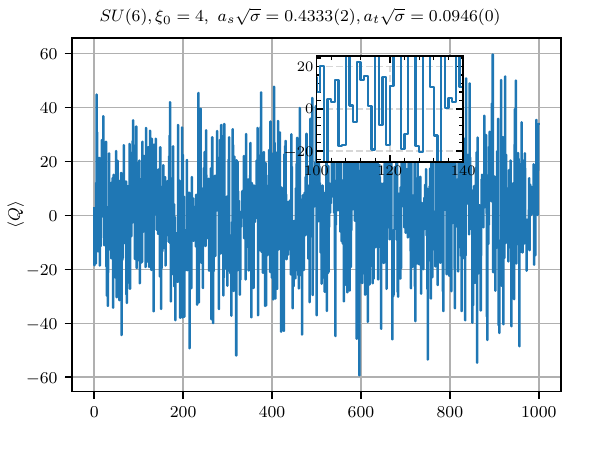}
    \end{minipage}
    \caption{The history of the topological charge for the ensembles produced for $N_c=3$ and $N_c=6$ for both anisotropies.}
    \label{fig:top_charge}
\end{figure*}

\subsection{Topological Charge and Critical Slowing down}
\label{sec:topology}

 The topological charge $Q$ of a gauge field is formally defined in the continuum as the four-dimensional Euclidean integral of the topological density across space-time. Its expression is given by:  
\begin{eqnarray}
	Q = \frac{1}{32\pi^2} \int d^4 x \: \epsilon_{\mu\nu\rho\sigma} \Tr\left[G_{\mu\nu}(x)G_{\rho\sigma}(x)\right] \,,
	\label{eq:Q_continuum_def}
	\end{eqnarray} 
where \( G_{\mu\nu}(x) \) is the field strength tensor, \( \epsilon_{\mu\nu\rho\sigma} \) is the Levi-Civita symbol in four dimensions, and \( \Tr \) denotes the trace over colour indices.  

On the lattice, any discretization method for the topological charge density that accurately reproduces the continuum expression in the limit of vanishing lattice spacing can be used. In this work, we employ the Clover representation \cite{Alexandrou:2017hqw} for the topological charge density.

In a Monte Carlo sequence of  $SU(N_c)$ lattice gauge fields, the rate at which the topological charge \( Q \) changes decreases significantly as the lattice spacing \( a(\beta) \) approaches zero at fixed \( N_c \), or as \( N_c \) increases while keeping \( a(\beta) \) constant. This phenomenon, is known as critical slowing down.
This rapid suppression of topological transitions raises important concerns about whether the generated gauge field configurations provide sufficient sampling across different topological sectors. In this work, we present a concise examination of the degree to which topological freezing influences our calculations and its potential implications for the reliability of our results. Consequently, the evaluation of the topological charge is utilized primarily as a test of the ergodicity of our simulations, rather than as a comprehensive study of its topological properties.

The extraction of the topological charge $Q$ out of a Monte-Carlo produced gauge field, is complicated by the existence of ultraviolet (UV) fluctuations in the gauge fields. These fluctuations obscure the topological structure of the field configuration, making direct extraction of $Q$ challenging. To address this issue, a smoothing procedure must be applied to suppress these UV divergences while preserving the essential topological features of the gauge fields. 

In our investigation, we adopt the APE smearing technique, which is both straightforward to implement and effective at reducing UV noise. APE smearing iteratively averages the gauge links to suppress high-momentum modes, thereby smoothing the gauge field. It is worth noting that at leading order, the topological content obtained through APE smearing is equivalent to other widely-used smoothing methods such as cooling, Stout smearing, and gradient flow, as demonstrated through gauge-field expansions. This equivalence has also been verified numerically in \( SU(3) \)~\cite{Alexandrou:2017hqw} gauge theories and can be generalized for $N_c>3$; the agreement of the results of the topological susceptibility with well-known established results directs to this equivalence. A detailed analysis of the topological quantities in anisotropic lattices for large-$N_c$ is underway.

Among these methods, gradient flow stands out for its well-established renormalizability properties, which ensure that the smoothing process does not introduce significant distortions to the physical content of the gauge field. However, implementing gradient flow is computationally more complex compared to APE smearing, especially for higher-dimensional gauge groups (\( N_c > 3 \)) or in simulations requiring large computational resources.

In summary, while gradient flow offers a theoretically robust framework for smoothing, APE smearing provides a practical alternative with sufficient accuracy for most applications, particularly when computational simplicity and efficiency are prioritized. 

In Fig.~\ref{fig:top_charge}, we present the evolution of the topological charge for both $SU(3)$ and $SU(6)$. The $SU(6)$ case represents our most extreme scenario, where severe topological freezing effects are expected. As seen in Fig.~\ref{fig:top_charge}, the topological charge fluctuates adequately for both $N_c=3$ and $N_c=6$, indicating that our simulations remain ergodic. While the charge evolution for $SU(6)$ exhibits less frequent changes, it does not compromise the diversity of sampled configurations across different topological charge sectors. Notably, we observe no signs of critical slowing down, which may be attributed to the chosen anisotropy. We also note that for larger values of $N_c$, we observed severe slowing down, which is why our calculations were limited up to $SU(6)$.

A natural question arises: how many smearing steps should be performed before quoting the topological charge? This can be addressed in two ways. One approach is to fix the number of smearing steps to match the corresponding gradient flow time, $t_0$~\cite{Alexandrou:2017hqw}. Alternatively, one can continue performing smearing until the topological susceptibility reaches a plateau, indicating that UV noise has been successfully removed while preserving the critical topological content of the configurations. In this study, we adopted the latter approach. Table~\ref{tab:ensemble} presents our computed values of the topological susceptibility. Comparing our values of the topological susceptibility with those reported in Ref.~\cite{Athenodorou:2021qvs}, which pertain to isotropic lattices, we find that they fall within the same bulk range. As a reference, an indicative value for the continuum limit of pure \( SU(3) \) gauge theory is \(\chi^{1/4}/\sqrt{\sigma} = 0.4246(36)\).

A detailed analysis of the topological properties of anisotropic lattices in the large-$N_c$ limit is highly compelling. The enlarged lattice volume may enhance the system's ergodicity, making the study particularly promising. Work in this direction is currently ongoing.

\begin{figure*}[htbp]
    \begin{minipage}{\columnwidth}
        \includegraphics[scale=0.9]{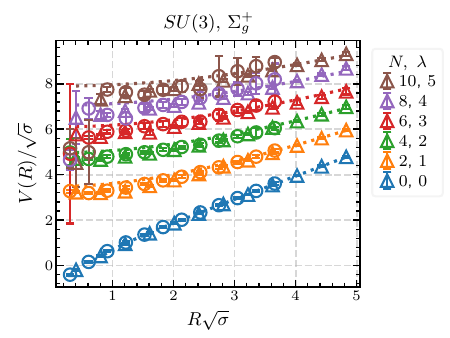}
    \end{minipage}\hfil
    \begin{minipage}{\columnwidth}
        \includegraphics[scale=0.9]{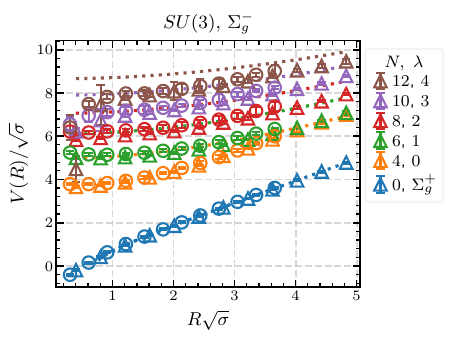}
    \end{minipage}\vfil
    \begin{minipage}{\columnwidth}
        \includegraphics[scale=0.9]{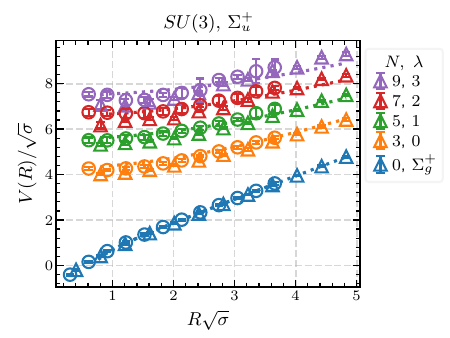}
    \end{minipage}\hfil
    \begin{minipage}{\columnwidth}
        \includegraphics[scale=0.9]{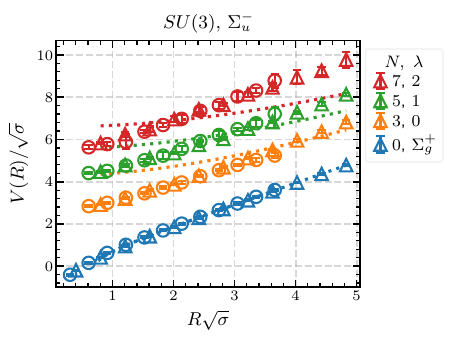}
    \end{minipage}\vfil
    \begin{minipage}{\columnwidth}
        \includegraphics[scale=0.9]{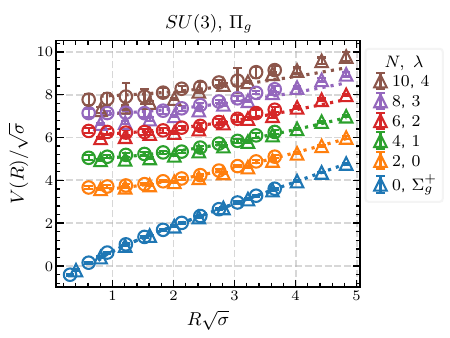}
    \end{minipage}\hfil
        \begin{minipage}{\columnwidth}
        \includegraphics[scale=0.9]{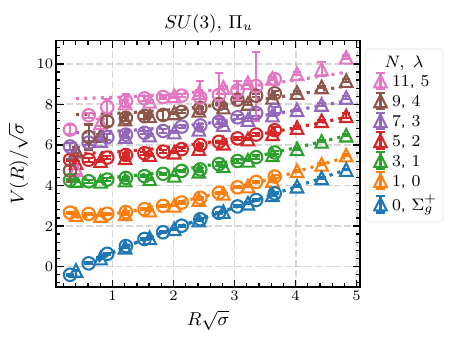}
    \end{minipage}\vfil
       \begin{minipage}{\columnwidth}
        \includegraphics[scale=0.9]{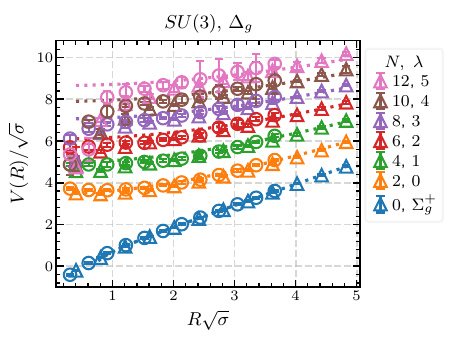}
    \end{minipage}\hfil
    \begin{minipage}{\columnwidth}
        \includegraphics[scale=0.9]{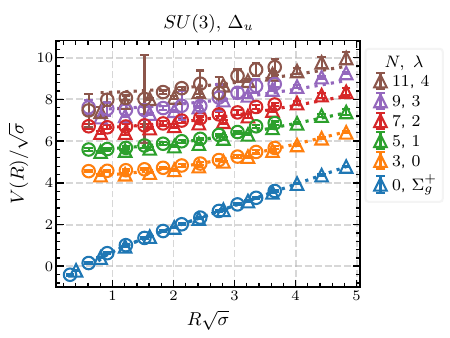}
    \end{minipage}
    \vspace{0.00cm}
    \caption{The spectrum for $SU(3)$ and the eight irreducible representations  \( \Sigma^+_g \), \( \Sigma^-_g \), \( \Sigma^+_u \), \( \Sigma^-_u \), \( \Pi_g \), \( \Pi_u \), \( \Delta_g \), and \( \Delta_u \) as these appear above each individual plot. Polygonal shapes in each figure denote the results obtained from the simulation with $\xi=4$, while circle markers indicate those for $\xi=2$. $N$ is the quantum number as this appears in Eq.~\ref{eq:Nambu_Goto} and defined in Eq.~\ref{eq:occupation_number}, and $\lambda$ is the radial excitation number. The dashed lines represent the Nambu-Goto predictions, with energy levels corresponding to the value of $N$, indicated by the matching colors in the legend on the right.\label{fig:SU3spectra}}
\end{figure*}

\begin{figure*}[htbp]

    \begin{minipage}{\columnwidth}
        \includegraphics[scale=0.9]{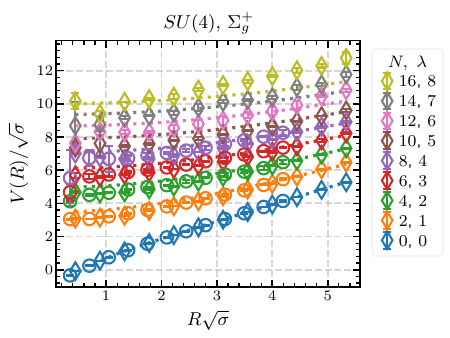}
    \end{minipage}\hfil
    \begin{minipage}{\columnwidth}
        \includegraphics[scale=0.9]{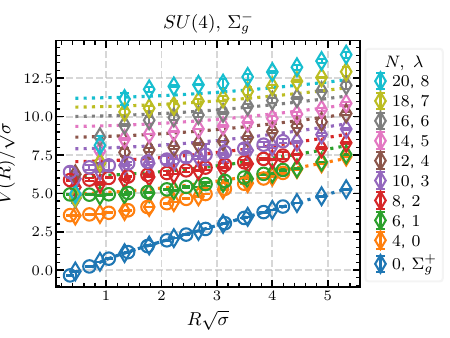}
    \end{minipage}\vfil
    \begin{minipage}{\columnwidth}
        \includegraphics[scale=0.9]{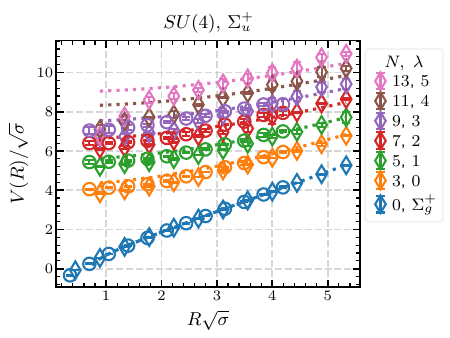}
    \end{minipage}\hfil
    \begin{minipage}{\columnwidth}
        \includegraphics[scale=0.9]{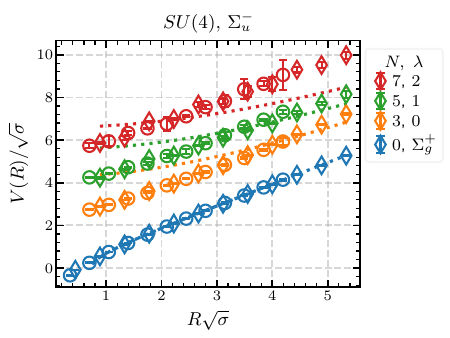}
    \end{minipage}\vfil
    \begin{minipage}{\columnwidth}
        \includegraphics[scale=0.9]{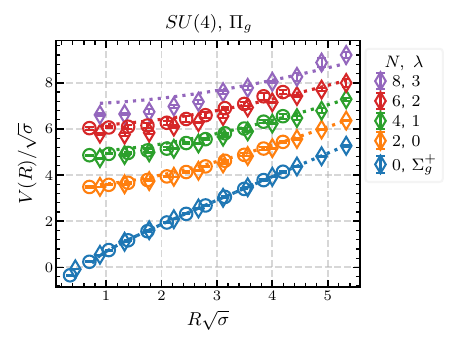}
    \end{minipage}\hfil
        \begin{minipage}{\columnwidth}
        \includegraphics[scale=0.9]{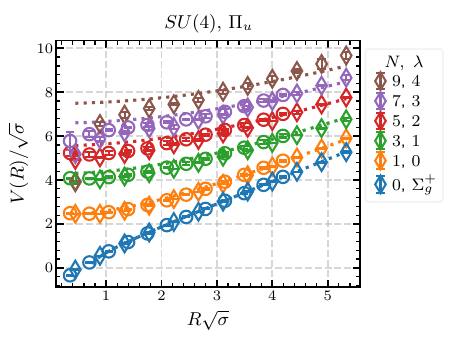}
    \end{minipage}\vfil
       \begin{minipage}{\columnwidth}
        \includegraphics[scale=0.9]{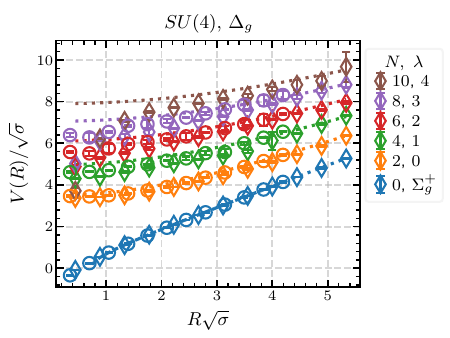}
    \end{minipage}\hfil
    \begin{minipage}{\columnwidth}
        \includegraphics[scale=0.9]{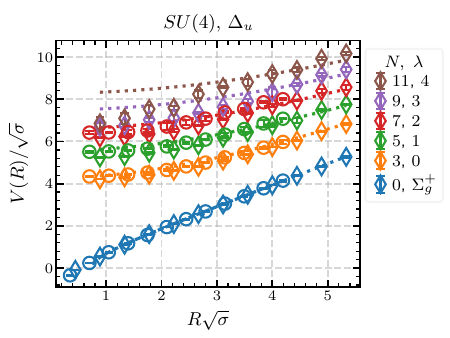}
    \end{minipage}
    \caption{The spectrum for $SU(4)$ and the eight irreducible representations  \( \Sigma^+_g \), \( \Sigma^-_g \), \( \Sigma^+_u \), \( \Sigma^-_u \), \( \Pi_g \), \( \Pi_u \), \( \Delta_g \), and \( \Delta_u \) as these appear above each individual plot. Polygonal shapes in each figure denote the results obtained from the simulation with $\xi=4$, while circle markers indicate those for $\xi=2$. $N$ is the quantum number as this appears in Eq.~\ref{eq:Nambu_Goto} and defined in Eq.~\ref{eq:occupation_number}, and $\lambda$ is the radial excitation number. The dashed lines represent the Nambu-Goto predictions, with energy levels corresponding to the value of $N$, indicated by the matching colors in the legend on the right.    \label{fig:SU4spectra}}
\end{figure*}

\begin{figure*}[htbp]
	\centering
    \begin{minipage}{\columnwidth}
        \includegraphics[scale=0.9]{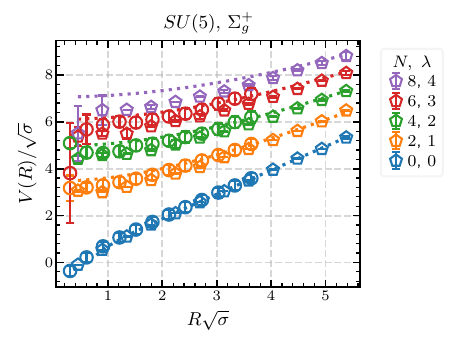}
    \end{minipage}\hfil
    \begin{minipage}{\columnwidth}
        \includegraphics[scale=0.9]{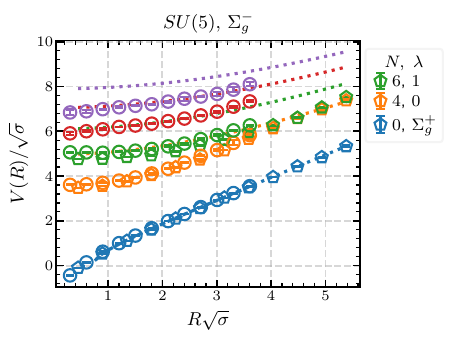}
    \end{minipage}\vfil
	\begin{minipage}{\columnwidth}
	\includegraphics[scale=0.9]{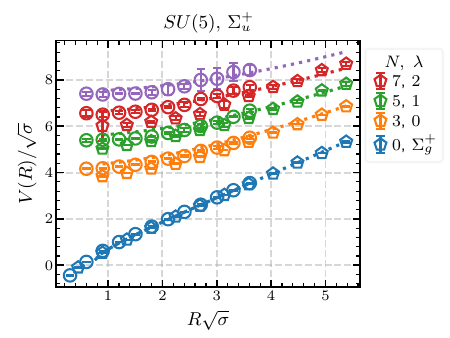}
      \end{minipage}\hfil
    \begin{minipage}{\columnwidth}
	\includegraphics[scale=0.9]{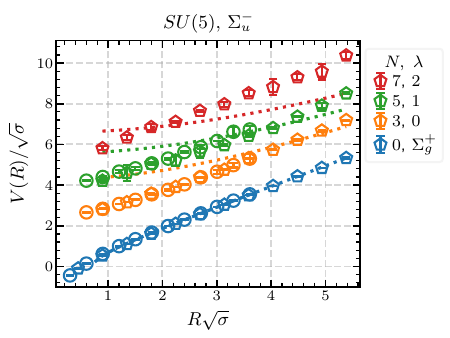}
      \end{minipage}\vfil
    \begin{minipage}{\columnwidth}
	\includegraphics[scale=0.9]{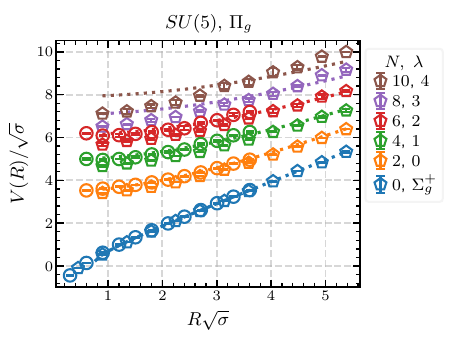}
      \end{minipage}\hfil
    \begin{minipage}{\columnwidth}
	\includegraphics[scale=0.9]{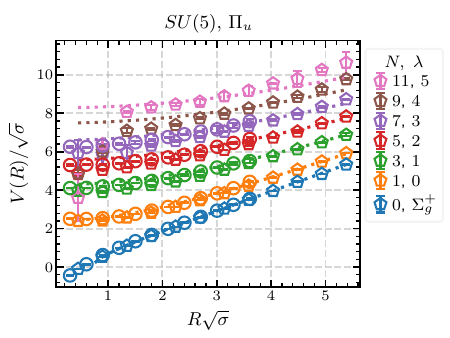}
      \end{minipage}\vfil
    \begin{minipage}{\columnwidth}
	\includegraphics[scale=0.9]{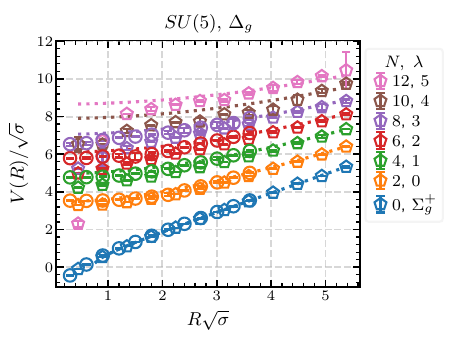} 
      \end{minipage}\hfil
    \begin{minipage}{\columnwidth}
	\includegraphics[scale=0.9]{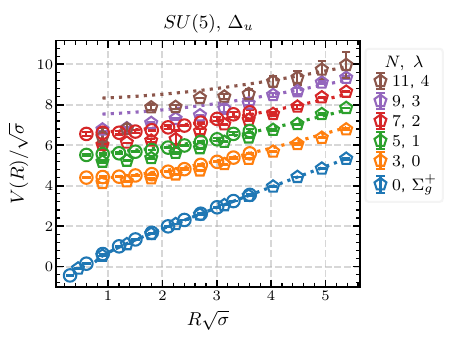}
      \end{minipage}
\caption{The spectrum for $SU(5)$ and the eight irreducible representations  \( \Sigma^+_g \), \( \Sigma^-_g \), \( \Sigma^+_u \), \( \Sigma^-_u \), \( \Pi_g \), \( \Pi_u \), \( \Delta_g \), and \( \Delta_u \) as these appear above each individual plot. Polygonal shapes in each figure denote the results obtained from the simulation with $\xi=4$, while circle markers indicate those for $\xi=2$. $N$ is the quantum number as this appears in Eq. \ref{eq:Nambu_Goto} and defined in~Eq.~\ref{eq:occupation_number}, and $\lambda$ is the radial excitation number. The dashed lines represent the Nambu-Goto predictions, with energy levels corresponding to the value of $N$, indicated by the matching colors in the legend on the right.   \label{fig:SU5spectra}}
\end{figure*}

\begin{figure*}[htbp]
	\centering
    \begin{minipage}{\columnwidth}
	\includegraphics[scale=0.9]{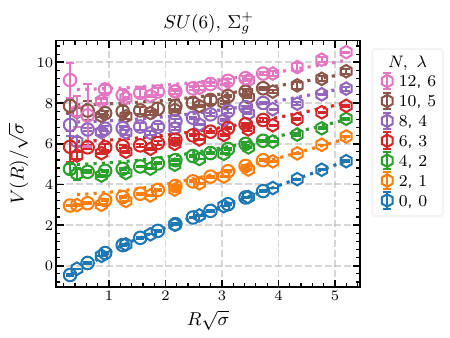}
    \end{minipage}\hfil
    \begin{minipage}{\columnwidth}
	\includegraphics[scale=0.9]{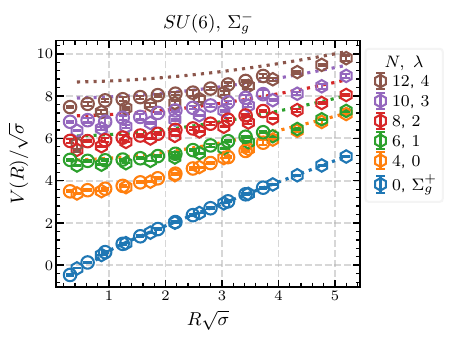}
    \end{minipage}\vfil
    \begin{minipage}{\columnwidth}
	\includegraphics[scale=0.9]{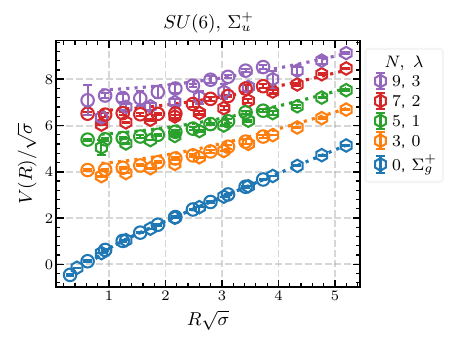}
    \end{minipage}\hfil
    \begin{minipage}{\columnwidth}
	\includegraphics[scale=0.9]{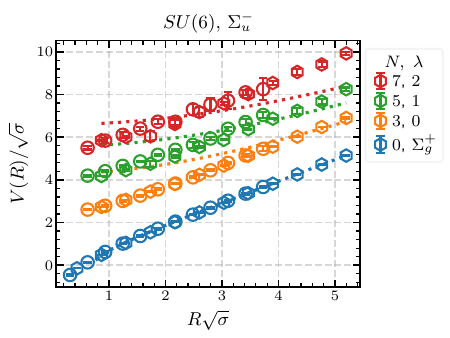}
    \end{minipage}\vfil
    \begin{minipage}{\columnwidth}
	\includegraphics[scale=0.9]{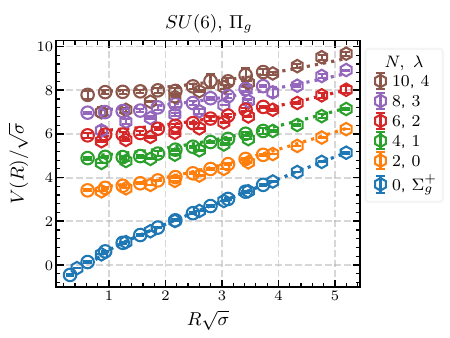}
    \end{minipage}\hfil
    \begin{minipage}{\columnwidth}
	\includegraphics[scale=0.9]{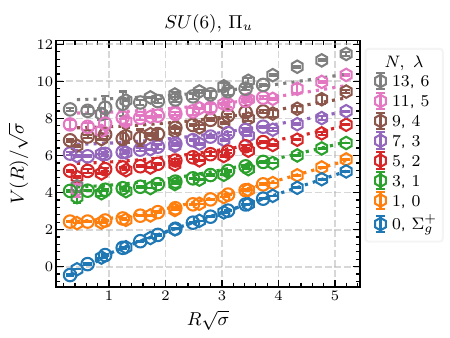}
    \end{minipage}\vfil
    \begin{minipage}{\columnwidth}
	\includegraphics[scale=0.9]{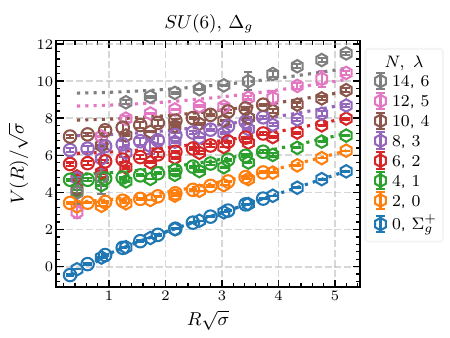} 
    \end{minipage}\hfil
    \begin{minipage}{\columnwidth}
	\includegraphics[scale=0.9]{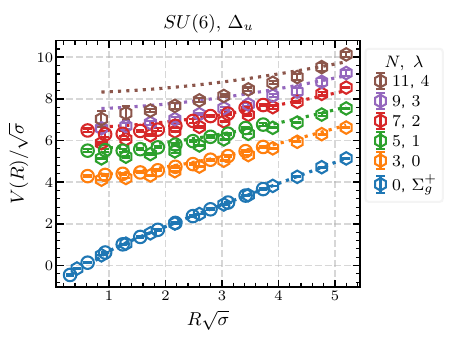}
    \end{minipage}
\caption{The spectrum for $SU(6)$ and the eight irreducible representations  \( \Sigma^+_g \), \( \Sigma^-_g \), \( \Sigma^+_u \), \( \Sigma^-_u \), \( \Pi_g \), \( \Pi_u \), \( \Delta_g \), and \( \Delta_u \) as these appear above each individual plot. Polygonal shapes in each figure denote the results obtained from the simulation with $\xi=4$, while circle markers indicate those for $\xi=2$. $N$ is the quantum number as this appears in Eq. \ref{eq:Nambu_Goto} and defined in Eq.~\ref{eq:occupation_number}, and $\lambda$ is the radial excitation number.  The dashed lines represent the Nambu-Goto predictions, with energy levels corresponding to the value of $N$, indicated by the matching colors in the legend on the right \label{fig:SU6spectra}}
\end{figure*}

\color{black}

 \section{Results on the spectrum of the open flux-tube}
\label{sec:results}

In this section, we present our results for the spectra obtained for \( N_c = 3 \) to \( N_c = 6 \), across the eight irreducible representations: \( \Sigma^+_g \), \( \Sigma^-_g \), \( \Sigma^+_u \), \( \Sigma^-_u \), \( \Pi_g \), \( \Pi_u \), \( \Delta_g \), and \( \Delta_u \), for both values of the bare anisotropy. Due to the richness of the spectral data, we choose to display the results in separate plots for each irreducible representation, allowing for clearer comparisons. The spectra are shown in Figs.~\ref{fig:SU3spectra}, \ref{fig:SU4spectra}, \ref{fig:SU5spectra}, and \ref{fig:SU6spectra}, corresponding to \( N_c = 3, 4, 5 \), and 6, respectively. Each plot includes data for both values of the bare anisotropy, 2 and 4. We emphasize once again that a higher anisotropy corresponds to a coarser lattice. For each $N_c$, we choose to present results at two different lattice spacings to examine whether the spectrum’s behaviour changes as we approach the continuum limit. Therefore, rather than attempting a continuum extrapolation, we perform a straightforward comparison to investigate the qualitative structure of the spectrum.

\begin{figure*}[t!]
\begin{minipage}{\columnwidth}
      \includegraphics[scale=1]{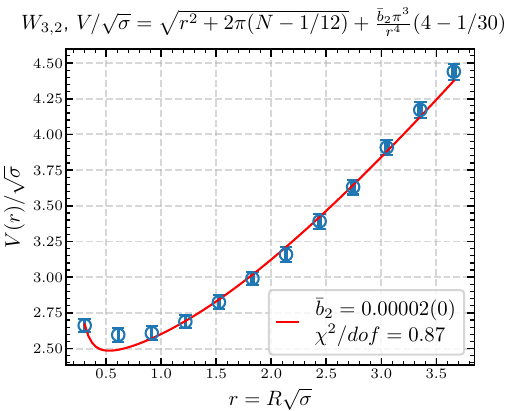}
\end{minipage}\hfil
\begin{minipage}{\columnwidth}
      \includegraphics[scale=1]{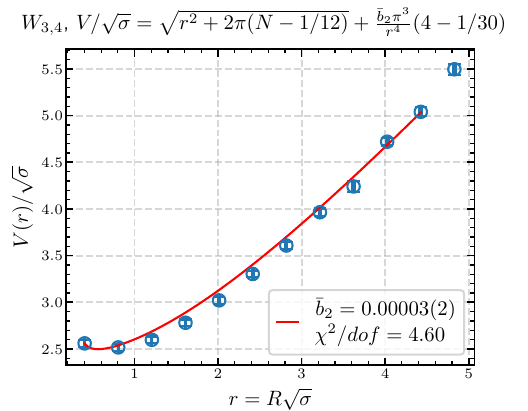}
\end{minipage}
      
    \caption{Searching for effective string theory parameters in the $1/ R^4$ term of the groundstate of the $\Pi_u$ spectrum, for our different anisotropies and lattice spacings.
\label{fig:1overR4}}
\end{figure*}

As explained in several sections throughout this manuscript, the primary focus of this work is the extraction and comparison of the spectra with the Nambu-Goto string prediction, specifically the Arvis potential at different excitation levels, as a means to investigate potential axion signatures. In addition, we aim to explore whether the spectrum exhibits any alterations as we approach the large-$N_c$ limit.

\subsubsection{The ground and first excited states}

We begin the investigation, with the absolute ground state which corresponds to the $\Sigma^+_g$ irreducible representation. The ground state can be interpreted as the vacuum state $|0\rangle$, with no phonon operators acting on it, implying that no scattering processes occur along the string’s worldsheet. Thus, we expect the spectrum to be well-approached by the Nambu-Goto model up to boundary terms setting in at order of $1/(R\sqrt{\sigma})^4$. We present results for the $\Sigma^+_g$ in all the plots appear in Figs.~\ref{fig:SU3spectra}, \ref{fig:SU4spectra}, \ref{fig:SU5spectra}, and \ref{fig:SU6spectra}, mainly for comparison purposes for higher excitation levels. Indeed, the $\Sigma^+_g$ groundstate is closely approximated by the Arvis potential, expanded just up to the Coulomb order $O(1/(R\sqrt{\sigma}))$ to remove the tachyonic divergence at short distances. Deviations of the leading order $O(1/(R\sqrt{\sigma})^4)$, as derived in \cite{Aharony:2009gg}, are expected to appear for finite values of $N_c$~\cite{Brandt:2017yzw} as well as in large-$N_c$ studies~\cite{Gonzalez-Arroyo:2012euf}, since these are $N_c$ independent. However, the focus of this work is not on subleading corrections in each state's energy but rather on comparing the spectra with the Nambu-Goto string to investigate potential massive excitation signatures, or effects of order $O(1/(R\sqrt{\sigma})^0)$. For simplicity, we will henceforth refer to such massive excitations as 'axions', in light of recent findings related to the closed flux tube ~\cite{Athenodorou:2024loq}. States containing axionic degrees of freedom are expected to behave as the ground state plus a constant massive excitation.

The first excited state in the spectra corresponds to the ground state of the $\Pi_u$ representation, shown in orange in the plots in Figs.~\ref{fig:SU3spectra}, \ref{fig:SU4spectra}, \ref{fig:SU5spectra}, and \ref{fig:SU6spectra}. In each case, it appears in the third row from the top and second column. Notably, it shows little sensitivity to finite lattice spacing effects, suggesting that the observed behaviour reflects continuum physics. This state matches the Arvis prediction with $N=1$ to good accuracy, indicating that it carries a single phonon with one unit of momentum.  This is expected, given that scattering processes along the world-sheet are absent in this case. In terms of phonon creation and annihilation operators, this state is described by $\alpha^{\dagger}_{\pm}|0\rangle$~\cite{Juge:2003ge}.

\begin{figure*}[t!]
  \begin{minipage}{\columnwidth}
    \hspace{-4cm}
	\includegraphics[scale=0.90]{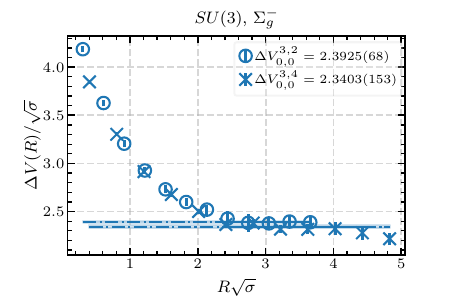}
    \end{minipage} \hspace{-2cm}
      \begin{minipage}{\columnwidth}
	\includegraphics[scale=0.90]{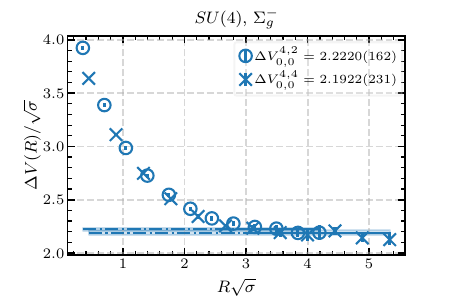}
    \end{minipage}\vfil
    \hspace{-2cm}
      \begin{minipage}{\columnwidth}
	\includegraphics[scale=0.90]{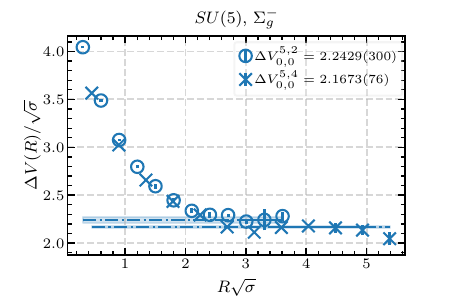}
      \end{minipage}\hspace{0cm}
      \begin{minipage}{\columnwidth}
	\includegraphics[scale=0.90]{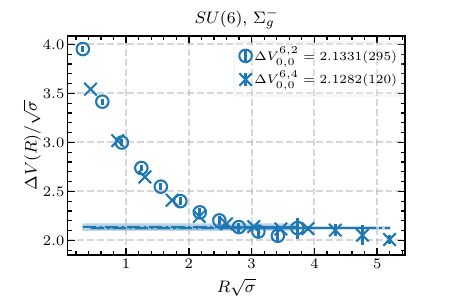}
      \end{minipage}\vfill
      \begin{minipage}{\columnwidth}
	\includegraphics[scale=0.90]{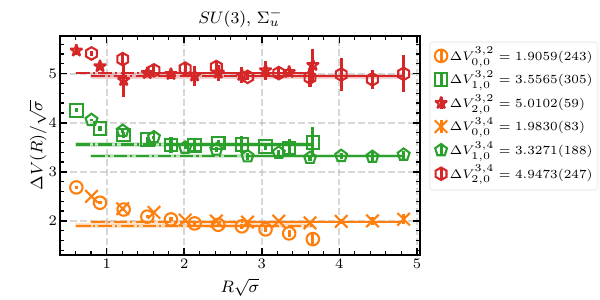}
      \end{minipage}\hfill
      \begin{minipage}{\columnwidth}
	\includegraphics[scale=0.90]{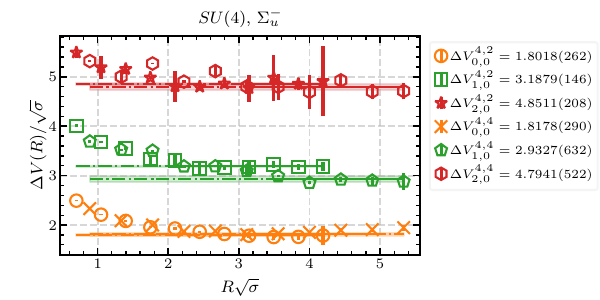}
      \end{minipage}\vfill
      \begin{minipage}{\columnwidth}
	\includegraphics[scale=0.90]{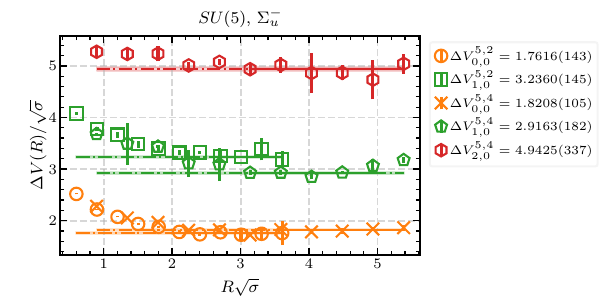}
      \end{minipage}\hfill
      \begin{minipage}{\columnwidth}
	\includegraphics[scale=0.90]{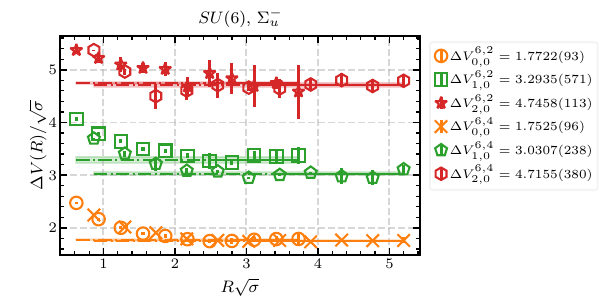}
      \end{minipage}
\caption{In the above plots we subtract the ground state potential to find evidence for massive resonances, for different gauge groups $SU(N_c)$ as well as for the two irreducible representations ${\Sigma_g^-}$ and ${\Sigma_u^-}$. We define $\Delta V^{N_c, \xi}_{\lambda, 0}= V_{{\Sigma_g^-}^\lambda}-V_{\Sigma_g^+}$ and $V_{{\Sigma_u^-}^\lambda}-V_{\Sigma_g^+}$ while the indices are as follows: $N_c$ is the number of colours, $\xi$ is the bare anisotropy, and $\lambda$ is the radial excitation number.
\label{fig:axionplateau}}
\end{figure*}

For the first two energy levels, where Nambu-Goto predictions provide a good description and no phonon interactions occur along the world-sheet, we may be able to observe effects arising from boundary terms—provided the numerical accuracy is sufficient to resolve them. We begin by analysing deviations from the Nambu-Goto model at order $O(1/(R\sqrt{\sigma})^4)$, as derived in~\cite{Aharony:2009gg}, Eq.~(\ref{eq:Nambu_goto_b2}), and anticipated in large-$N_c$ studies~\cite{Gonzalez-Arroyo:2012euf}. 

To this end, we fit our zero-phonon and one-phonon potentials—corresponding to the ground states of the $\Sigma_g^+$ and $\Pi_u$ channels, respectively—with the Nambu-Goto expression plus the additional boundary contributions predicted by effective string theory. We note that $B_0 = 0$ and $B_1 = 4$ for the $\Sigma_g^+$ and $\Pi_u$ ground states, respectively.

In the case of the $\Sigma_g^+$ channel, deviations from Nambu-Goto are negligible, and due to the limits of our numerical precision, the coefficient $\bar{b}_2$ could not be extracted with meaningful accuracy. Consequently, the most promising candidate for identifying such deviations is the first excited Arvis state, corresponding to the $\Pi_u$ channel. In Fig.~\ref{fig:1overR4}, we present the fits for this case. As the fit results indicate, the extracted value of $\bar{b}_2$ is close to zero, and within our current precision, we are unable to claim a statistically significant deviation from the Nambu-Goto model.

A definitive detection of such deviations would require an exponentially larger statistical sample, which could be achieved through the use of multi-level algorithms. However, such an investigation lies beyond the scope of the present work.

Calculations of $\bar{b}_2$ have been performed in $D=2+1$ dimensions for various gauge groups, including $Z_2$, $U(1)$, and $SU(N_c)$ with $N_c=2,3,\dots,6$~\cite{Brandt:2010bw,Brandt:2013eua,Brandt:2016xsp,Brandt:2018fft,Billo:2012da,Caselle:2024ent,Caselle:2024zoh,Baffigo:2023rin}. However, due to the considerable numerical effort, an unambiguous determination in $D=3+1$ dimensions remains absent from the literature.

\subsubsection{The $N=2$ excited states}

We now turn our attention to the study of higher excited states of the open flux-tube. A general observation is that deviations from the Nambu-Goto predictions become more pronounced as the energy increases.

The next relevant string excitation corresponds to \( N~=~2 \). We expect to encounter such states in the following irreducible representations: as the first excited state of \( \Sigma_g^+ \), and as the ground states of \( \Pi_g \) and \( \Delta_g \). This expectation is indeed confirmed by our spectrum. Moreover, we observe that these states exhibit only small deviations from the Nambu-Goto predictions. We can confidently state that these three states are purely string-like, with no indications of axionic contributions in the spectrum.

\subsubsection{The $N=3$, $N=4$ excited states and the axion}

We now turn to the next string excitation corresponding to \( N = 3 \). According to Table~\ref{table:quantum_numbers}, this energy level is expected to be six-fold degenerate. The first irreducible representation in which we expect to encounter such a state is \( \Sigma_{u}^{+} \). Indeed, we observe a state in this channel that shows minor deviations from the Nambu--Goto prediction, thereby confirming its string-like nature.

Next, we examine the first excited state in the \( \Pi_{u} \) representation. This state exhibits good agreement with the Arvis potential, with only small deviations. According to Table~\ref{table:quantum_numbers}, we would expect to find another energy level near the \( N = 3 \) string prediction, manifesting as the second excited state in the \( \Pi_{u} \) representation. Instead, we find that the next excitation aligns more closely with the \( N = 5 \) prediction from the Arvis potential, which is somewhat surprising. Of course, we must bear in mind that this is a variational calculation, and the current operator basis may lack the necessary features to project effectively onto the expected state. Nevertheless, the low-lying spectrum in the \( \Pi_{u} \) representation appears to be composed exclusively of string-like states.

Turning now to the \( \Delta_g \) irreducible representation, we observe that its ground state is also in good agreement with the Arvis potential, again with minor deviations indicative of string-like behaviour.

Finally, we consider the \( \Sigma_{u}^{-} \) representation, which carries the same quantum numbers as the axion in the closed flux-tube spectrum. Surprisingly, not only the ground state but the entire spectrum in this channel deviates significantly from the Nambu--Goto prediction. These states resemble ground states with an additional constant mass term, as illustrated in the plots shown in the second column and second row of Figs.~\ref{fig:SU3spectra}, \ref{fig:SU4spectra}, \ref{fig:SU5spectra}, and \ref{fig:SU6spectra}. As \( N_c \) increases, these states exhibit negligible dependence on lattice artifacts, and the qualitative features of the spectrum remain unchanged. This behaviour suggests that the deviation is a genuine large-\( N_c \) effect, reflecting an intrinsic property of the world-sheet theory. 

We now turn to the \( N = 4 \) excitation level, which is expected to be twelve-fold degenerate. As shown in Figs.~\ref{fig:SU3spectra}, \ref{fig:SU4spectra}, \ref{fig:SU5spectra}, and \ref{fig:SU6spectra}, the states in all accessible irreducible representations appear to exhibit string-like behaviour, with the notable exception of the \( \Sigma_{g}^{-} \) channel. In this representation, the ground state shows significant deviations from the Nambu--Goto prediction, while the first excited state appears to align with the ground-state level of the Arvis potential rather than its first excitation. This behaviour suggests the presence of non-string-like features in the \( \Sigma_{g}^{-} \) spectrum. This behaviour remains unaltered as we move to the large-$N_c$ limit.

\begin{figure*}[t!]
\begin{minipage}{\columnwidth}
            \includegraphics[scale=0.9]{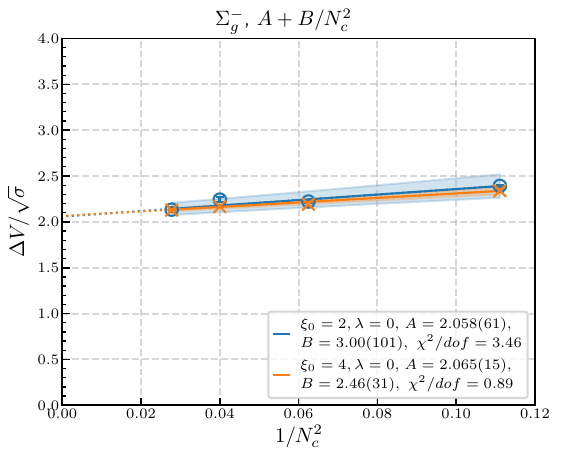}
\end{minipage}
\begin{minipage}{\columnwidth}
            \includegraphics[scale=0.9]{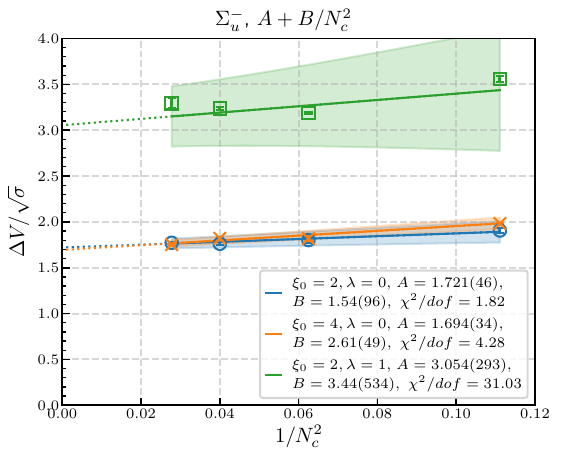}
\end{minipage}\vfil
    \caption{Comparing axion's masses for different ensembles and their extrapolations as a function of $1/N_c^2$.
\label{fig:axionmasses}}
\end{figure*}

The observed behaviour is consistent with expectations for a massive excitation coupled to the string. In Fig.~\ref{fig:axionplateau}, we illustrate how we extract the axion mass for $N_c = 3, 4, 5, 6$. Specifically, we determine a constant mass shift relative to the absolute ground state of the $\Sigma_g^+$ channel in the large-distance limit ($R \to \infty$). To this end, we subtract the absolute ground state energy from the energy level of interest and plot the resulting mass shift as a function of $R$ to identify a plateau.

We find evidence of a state with axion-like characteristics in the $\Sigma_g^-$ spectrum, and three such states in the $\Sigma_u^-$ spectrum. However, the third excited state in the $\Sigma_u^-$ channel (with $\lambda = 2$) does not exhibit a clear plateau at the largest value of $N_c$, suggesting that it may be subject to significant systematic effects due to its large mass. Nevertheless, for the remaining states, the presence of well-defined plateaus supports the interpretation of constant-mass excitations propagating along the worldsheet of the flux-tube.

As discussed above, when comparing the plots for the $\Sigma_{g}^{-}$ and $\Sigma_{u}^{-}$ channels across different values of the number of colours, we observe that the masses of the corresponding resonances show only a weak dependence on $N_c$. However, this can be quantified, and we confirm that the resonance masses tend toward finite values in the large-$N_c$ limit. This behaviour can be captured through a large-$N_c$ expansion.

To this end, we perform extrapolations in $1/N_c^2$ for the ground state of the $\Sigma_{g}^{-}$ channel, and for both the ground and first excited states of the $\Sigma_{u}^{-}$ channel. The precision of the third excited state in $\Sigma_{u}^{-}$ is insufficient for meaningful statistical analysis. These extrapolations are carried out separately for each value of the anisotropy, under the assumption that the lattice spacings (in units of the string tension) are approximately equivalent. A fluctuation of order \( a_s \sqrt{\sigma} \sim 0.05 \) is present but has no significant impact on the extrapolations.

In Fig.~\ref{fig:axionmasses}, we show the large-$N_c$ extrapolations of the three well-defined axion-like masses, plotted as functions of $1/N_c^2$. Note that we observe slightly different masses for the two ensembles used for each $N_c$, which differ in anisotropy $\xi$, coupling $\beta$, and lattice spacing $a$.

The extrapolated axion masses we obtain are:
\begin{itemize}
    \item $m_{\rm axion} / \sqrt{\sigma} = 2.058(6)$ and $2.065(15)$ for $\xi=2$ and $4$ respectively and for the lightest $\Sigma_g^-$ state with $\lambda = 0$.
    \item $m_{\rm axion}/ \sqrt{\sigma} = 1.721(46)$ and $1.694(34)$ for $\xi=2$ and $4$ respectively and for the lightest $\Sigma_u^-$ state with $\lambda = 0$.
    \item $m_{\rm axion} / \sqrt{\sigma} = 3.054(293)$ for $\xi=2$ and the first excited state $\Sigma_u^-$ state with $\lambda = 1$.
\end{itemize}

Moreover, the lightest axion mass is consistent with the value observed in the closed flux-tube spectrum~\cite{Athenodorou:2007du, Athenodorou:2010cs, Athenodorou:2011rx, Athenodorou:2024loq}. Intuitively, we expect these two states to carry similar quantum numbers. Although the open and closed flux-tube configurations differ in their characterization, one can argue that the $\Sigma_u^-$ quantum numbers, defined by $\Lambda = 0$, $\mathcal{C}o\mathcal{P} = -$, and $\epsilon = -$, are equivalent to the closed flux-tube assignment $J = 0$, $P_{||} = -$, and $P_{\perp} = -$, corresponding to $0^{--}$. Specifically, the quoted value reported in those works is $m_{\rm axion}/ \sqrt{\sigma} = 1.65(2)$ for both $SU(5)$ and $SU(6)$.


\section{Conclusion}
\vspace{-0.1cm}

In this study, we have conducted a comprehensive analysis of the open flux-tube spectrum in the large-$N_c$ limit.

Specifically, we extracted the spectra for $SU(3)$, $SU(4)$, $SU(5)$, and $SU(6)$ gauge theories, each at two different anisotropies and, thus, different lattice spacings.

Our results exhibit minimal finite lattice spacing effects and good convergence in the $N_c$ extrapolation, indicating that the observed physics closely approximates both the large-$N_c$ and continuum limits.

In addition, we examined the ergodicity properties of the generated ensembles by monitoring the topological charge histories on anisotropic lattices. Up to $N_c=6$, our simulations show evidence of ergodic behaviour.

Our findings reveal that most of the states in the flux-tube spectra at large colour-source separations can be well described by the Nambu-Goto string model, which leads to the Arvis potential, with minimal deviations. Within our framework, designed for the study of long flux-tubes, the $O(1/(R\sqrt{\sigma})^4)$ correction expected from the boundary terms in the effective string theories does not show up with significance.

However, several states for the irreducible representations $\Sigma_u^-$ and $\Sigma_g^-$ exhibit substantial deviations from Nambu-Goto behaviour, showing characteristics of massive excitations. These states persist even after extrapolation to the large-$N_c$ limit. Notably, the lightest of these massive modes, referred to as "axions", has a mass that matches the corresponding excitation observed in the closed flux-tube spectrum.

This study rules out the interpretation of these axions as glueballs coupled to the flux-tube, as they persist in the Large-$N_c$ limit, whereas hadron-hadron interactions are expected to vanish in this limit~\cite{Lucini:2012gg,Manohar:1998xv,tHooft:1973alw}.

The microscopic nature of these worldsheet axions, and their relation to QCD, remain unclear. They may originate from the phonon-phonon resonances that could be connected to the existence of a number of pseudoscalar fields on the worldsheet of the bosonic string. At this end we stress the need for the development of a TBA analysis~\cite{Athenodorou:2024loq} for the open flux-tube in the same manner this has been worked out for the case of the closed flux-tube. Another natural question arises: how many massive excitation fields, or axions, are present on the worldsheet of the flux-tube string? This could be answered within the context of a TBA analysis.

Interestingly, we observe that in the large-$N_c$ limit, the constituent components of hadrons—whether quarks~\cite{Bali:2013kia} or gluons~\cite{Lucini:2004my,Athenodorou:2021qvs}—persist, as glueball and meson masses tend toward finite values. This supports the possibility that the axion could be interpreted as a constituent gluon. Alternatively, it may be evidence for a longitudinal wave~\cite{Brower:2005pb}.

Several promising directions for future work include a more detailed study of the flux-tube at shorter distances, the implemention of a non-perturbative determination of the anisotropy factor, an extraction of the spectrum of flux-tubes in higher representations with $N$-ality $k$ ($k$-strings~\cite{Athenodorou:2016ndx,Athenodorou:2008cj}), a more refined characterization of the axion states using a broader operator basis, a systematic investigation of the topological effects in the spectrum and an exploration of the flux-tube spectra at finite temperatures.

\section*{Acknowledgments}
\vspace{-0.1cm}
	
The authors acknowledge discussions with Claudio Bonanno, Eric Braaten, Nuno Cardoso, Sergei Dubovsky, Victor Gorbenko, Roman Höllwieser, Mika Lauk, Biagio Lucini, Abhishek Mohapatra, Michael Teper and Marc Wagner.
AS was supported by CEFEMA UIDB/04540/2020 Post-Doctoral Research Fellowship. AA acknowledges support by the ``EuroCC" project funded by the ``Deputy Ministry of Research, Innovation and Digital Policy and the Cyprus Research and Innovation Foundation" as well as by the EuroHPC JU under grant agreement No 101101903. AS and PB thank CeFEMA, an IST research unit whose activities are partially funded by FCT contract  UIDB/04540 for R\&D Units. AA is also indebted to Conghuan Luo for performing a critical review of the manuscript.

\bibliographystyle{apsrev4-2}

\bibliography{references}

\end{document}